\documentclass[14pt]{extarticle}
\usepackage[margin=0.5in]{geometry}
\usepackage{amsmath}
\usepackage{amsfonts}
\usepackage{graphicx}
\usepackage{floatrow}
\usepackage{authblk}
\usepackage[sort&compress,numbers]{natbib}

\newcommand{\textover}[2]{\hphantom{#2}\llap{$\mathsurround=0pt{#1}$}}
\newcommand{\isep}{\mathrel{{.}\,{.}}\nobreak}

\providecommand{\keywords}[1]
{   \small
	\textbf{\textit{Keywords---}} #1
}

\begin{document}

\title{Large-amplitude periodic atomic vibrations in diamond}

\author{George Chechin\footnote{gchechin@gmail.com}}
\author{Denis Ryabov}
\author{Stepan Shcherbinin}
\affil{Southern Federal University, Research Institute of Physics, Stachki Ave.\ 194, Rostov-on-Don, 344090, Russia}
\date{}
\maketitle

\begin{abstract}
Symmetry-determined nonlinear normal modes, which describe large-amplitude vibrations of diamond lattice, are studied by specific group-theoretical methods. For two of these modes, corresponding to vibrational state with and without multiplication of the unit cell, we present their atomic patterns and dynamical properties obtained with the aid of the ab initio simulations based on the density functional theory.
\end{abstract}

\section{Introduction}

Most of the papers, devoted to the study of atomic vibrations in 3D crystals, are based on harmonic approximation, which permits introducing conventional (linear) normal modes. These modes are independent of each other and their quantization leads to the concept of phonons, which is of great importance in the physics of crystals. However, linear normal modes are no longer exact solutions of nonlinear dynamical equations beyond the harmonic approximation. In \cite{Ros} was introduced the concept of nonlinear normal modes, which are a generalization of conventional normal modes. Unfortunately, the Rosenberg modes, which are exact solutions of nonlinear differential equations, can exist only in very special mechanical systems.

On the other hand, in the papers \cite{DAN-1, PhysD-98} the concept of \textit{bushes of nonlinear normal modes} for arbitrary physical systems with \textit{discrete symmetry} was introduced. These dynamical objects can be found with the aid of specific group-theoretical methods and they represent \textit{exact} solutions of nonlinear equations, regardless of the type of interparticle interactions in the considered system. Every Rosenberg mode describes a periodic vibration, while a given bush determines quasi-periodical vibration whose number of basis frequencies is equal to the number of nonlinear normal modes entering into the bush.

In our previous works \cite{FPU-1, FPU-2, octo, Zhukov} we investigated the bushes of vibrational nonlinear normal modes in various model systems described by different groups of point and space symmetry. First-principle simulations based on the density functional theory (DFT) allow one to investigate bushes of modes in the framework of much more realistic models of physical systems. With the aid of such approach, we have investigated low-dimensional vibrational bushes in graphene \cite{graphene-1, graphene-2}, carbyne \cite{carbyne}, and the molecule $SF_6$ \cite{SF6}.

In the present paper we apply group-theoretical methods, as well as DFT-simulations to study large-amplitude atomic vibrations in diamond, which correspond to nonlinear normal modes (one-dimensional bushes).

\section{The concept of nonlinear normal modes}

There exist two different concepts of nonlinear normal modes (NNMs): the modes by Lyapunov and the modes by Rosenberg. Both types of these dynamical objects correspond to certain periodic vibrations in nonlinear physical systems. Lyapunov modes can be constructed as a continuation of linear normal modes with respect to a nonlinear parameter (existence of the harmonic approximation is assumed). As a result, the number of Lyapunov modes is equal to the number of degrees of freedom of the considered system. Unfortunately, existence domains of the Lyapunov modes turn out to be so small in practice, and the procedure for their construction is so cumbersome that these modes have not received wide application in solving physical problems.

In contrast, Rosenberg modes can exist for large values of nonlinearity parameter and they can be constructed with the aid of a relatively simple method. However, there is a significant limitation in practical use of Rosenberg modes because they can exist only in systems with very specific features. The most important class of mechanical systems that admit existence of the Rosenberg modes is formed by those systems whose potential energy is a homogeneous function of all its variables. From this point of view, Rosenberg modes seem to be exotic objects for studying real physical problems.

However, we proved in \cite{DAN-1, PhysD-98} that there are some symmetry-related causes for the dynamical systems with \textit{discrete symmetry} that dictate the possibility of existence of exact Rosenberg NNMs. Moreover, these modes can exist in such systems \textit{independent} on the type of interparticle interactions because solely the symmetry determines the possibility of their existence. This is the reason why we call them symmetry-determined Rosenberg nonlinear normal modes. Hereafter, only the modes of this type are discussed and we will omit the words ``symmetry-determined'' for brevity.

Let us consider Rosenberg NNMs in more detail. According to the definition, each Rosenberg NNM represents a periodical vibrational regime for which evolution of all $N$ degrees of freedom $x_i(t)$ are determined by the same time-dependent function $f(t)$:
\begin{equation}\label{RosNNMdef}
x_i(t)=a_i f(t), \quad i=1\isep N.
\end{equation}
Here $a_i$ represents amplitude of the oscillations of $i$\nobreakdash-th variable $x_i(t)$. Evidently, the definition~\eqref{RosNNMdef} is a generalization of that of the conventional (linear) normal modes that are exact solutions only in the harmonic approximation. In this case, $f(t)=\cos(\omega t+\varphi_0)$, where $\omega$ and $\varphi_0$ are frequency and initial phase, respectively.

The method for finding Rosenberg modes is reduced to substituting the anzats~\eqref{RosNNMdef} into the dynamical equations of the considered nonlinear system and demanding the equations, obtained in such a way, to be compatible with each other (as was already noted, this compatibility can only be achieved in some exceptional cases). Our group-theoretical methods for obtaining symmetry-determined Rosenberg modes are based on a completely different approach, which is considered in the next section.

\begin{figure}[h]
	\begin{minipage}[h]{0.49\linewidth}
		\center{\includegraphics[width=1\linewidth]{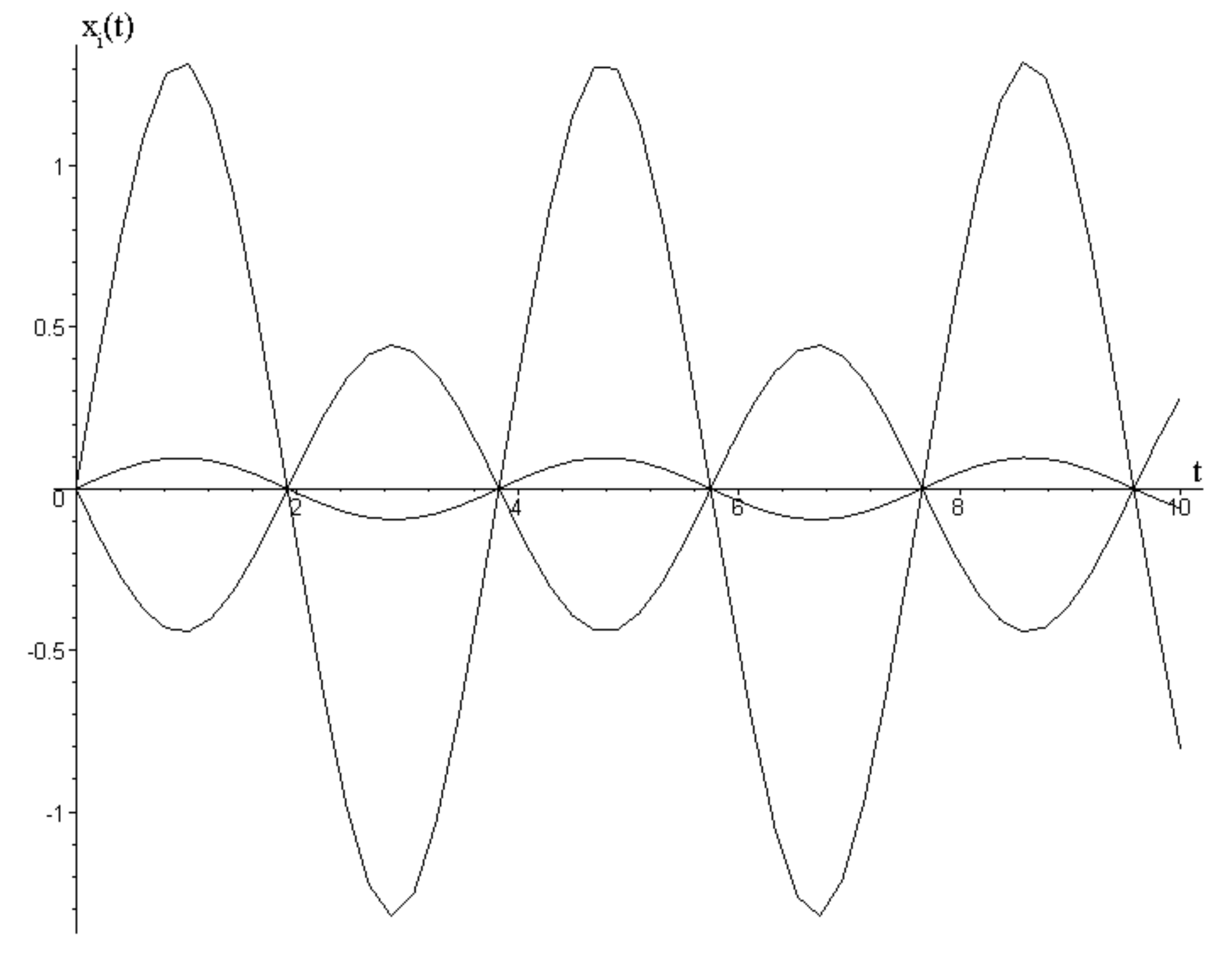} \\ (a)}
	\end{minipage}
	\hfill
	\begin{minipage}[h]{0.49\linewidth}
		\center{\includegraphics[width=1\linewidth]{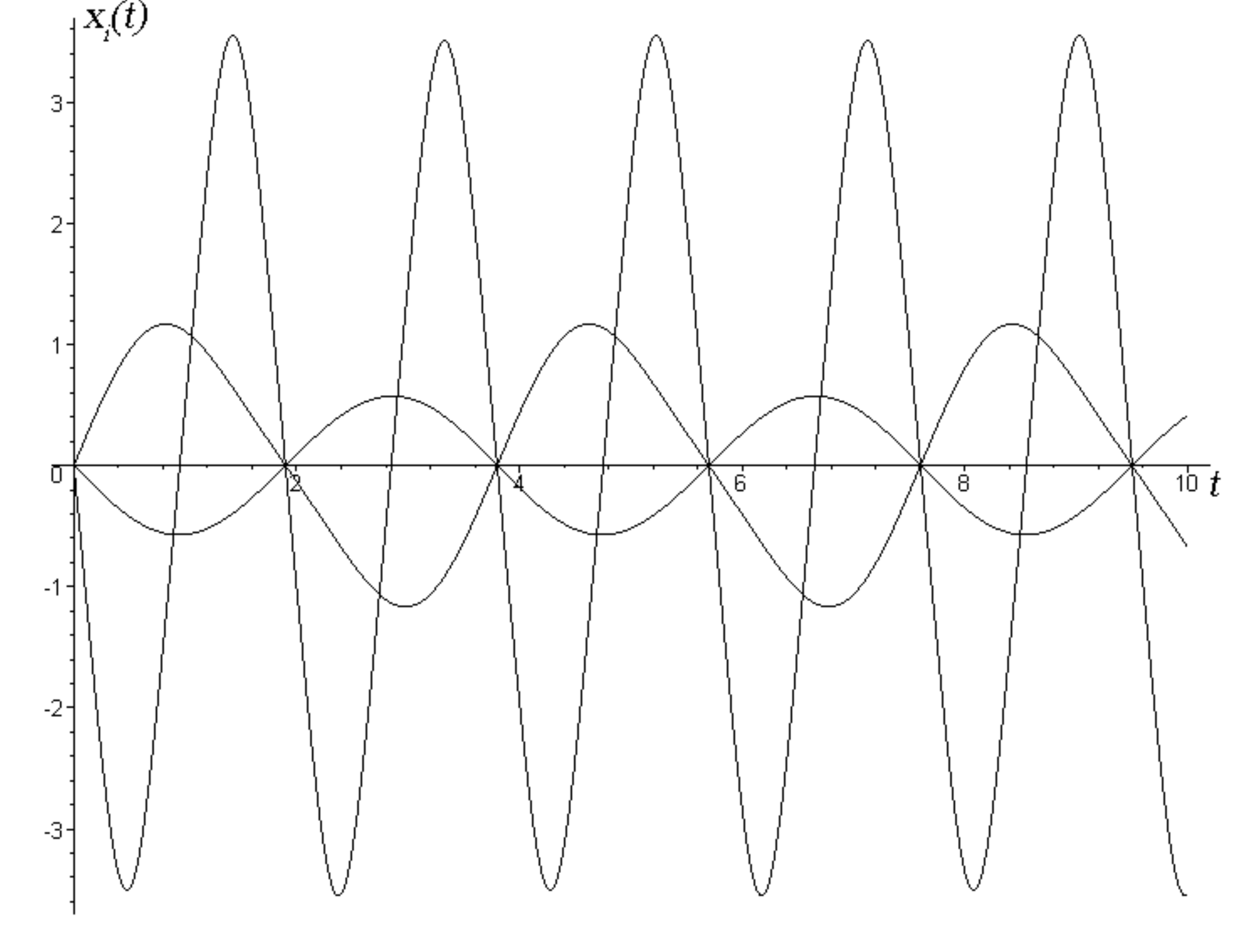} \\ (b)}
	\end{minipage}
	\caption{Examples of (a)~single-frequency and (b)~two-frequency discrete breathers.}
	\label{fig1}
\end{figure}

The Rosenberg NNM can be a localized, as well as a delocalized vibrational mode. If this mode is localized in the crystal lattice, it represents a \textit{discrete breather} (DB). The converse is not true, namely, not every discrete breather is a Rosenberg mode. We illustrate this fact in Fig.~\ref{fig1}, which demonstrates time evolution of displacements $x_i(t)$ for a chain whose particles interact via the homogeneous potential of the fourth order~\cite{SoundVibration}. Left plot corresponds to a true discrete breather, because all particles of the chain oscillate with the same frequency. In contrast, the right plot corresponds to \textit{multi-frequency} DB: the particles of the breather's tail oscillate with the frequency that is half the frequency of the particles of its core. As a result, when core particles pass through the equilibrium positions, the peripheral particles have non-zero displacements. This contradicts the definition of the Rosenberg mode~\eqref{RosNNMdef}, according to which the displacements of all particles of the chain must vanish simultaneously when $f(t)=0$.

However, the main objects studied in the present paper are delocalized Rosenberg NNMs, which tend to the linear normal modes for infinitesimal  amplitudes. Some relation of these modes to discrete breathers will be discussed in the Conclusion of the present paper.

\section{Bushes of nonlinear normal modes}

Let us consider the problem of finding some exact solutions of nonlinear dynamical equations for a given $N$\nobreakdash-particle Hamiltonian system. In the general case, we have no methods for solving this problem. However, if the considered system possesses some discrete symmetry group, there exists a fruitful idea which is realized in the theory of bushes of nonlinear normal modes. As the first step of this approach, we try to find some symmetry-related \textit{invariant manifolds}. This notion can be explained as follows.

The state of a given classical system is determined by the set of coordinates and momenta of all particles, i.e.\ by a certain point in its phase space. Since this point changes in time, the evolution of the considered system corresponds to a certain curve in the phase space. By definition, invariant manifold represents such set of points that if the system is on this manifold at the initial time, it will remain there forever, e.g., the entire trajectory of the system in phase space belongs to this manifold.

In general case, methods for finding invariant manifolds are unknown, but they do exist for the systems with discrete symmetry groups such as point and space groups. Under the term ``parent symmetry group'' $G_0$, we mean the symmetry group of the system's equilibrium state or symmetry group of its Hamiltonian\footnote{ If several equivalent equilibrium states are possible in the system, then the symmetry group of its Hamiltonian is higher than the symmetry groups of these states.}. A certain subgroup $G_j (j \neq 0)$ of the parent group $G_0$ can be ascribed to every possible vibrational regime in the considered system. Indeed, we can act on the displacement pattern $\textbf{X}(t)$, corresponding to this regime at arbitrary time, by all symmetry elements of the group $G_0$ and collect together those which do not change the pattern. The obtained collection represents the group $G_j$. Thus, we can say that above dynamical regime is invariant with respect to the subgroup $G_j$ of the group $G_0$:
\begin{equation}\label{InvCond}
\hat{G}_j\textbf{X}(t) = \textbf{X}(t).
\end{equation}
Here $\textbf{X}(t) = [x_1(t), x_2(t),\ldots,x_n(t)]$, while $\hat{G}_j$ is the group of operators acting on the vectors of $N$\nobreakdash-dimensional configuration space\footnote{Each operator $\hat{g}_i$ from the group $\hat{G}_j$ corresponds to the element $g_i$ of the symmetry group $G_j$ which acts in the conventional three-dimensional space.}.

On the contrary, if we choose a certain subgroup of the parent group $G_0$, we can find the corresponding displacement pattern, using the condition of invariance~\eqref{InvCond}. The vector $\textbf{X}(t)$ can be decomposed into a certain basis in the $N$\nobreakdash-dimensional configurational space ${\textbf{v}_1,\textbf{v}_2,\ldots,\textbf{v}_n}$:
\begin{equation}\label{ConfVectDecompose}
\textbf{X}(t)=\sum{c_i(t)\textbf{v}_i}
\end{equation}
The full collections of eigenvectors of the matrix of force constants can be used as the basis in~\eqref{ConfVectDecompose}. However, we use a more general and more convenient basis formed by basis vectors of all \textit{irreducible representations} (irreps) of the parent group $G_0$. These vectors must be constructed on the displacements of all atoms of the considered physical system, i.e.\ a specific displacement pattern corresponds to each of these vectors. It is very important, that the basis, chosen in such a way, does not depend on the specific type of interparticle interactions (unlike that constructed with the aid of the matrix of force constants). It can be obtained by specific group-theoretical methods~\cite{Chechin-Alone} using only the information about the symmetry and the geometrical structure of the system.

If we search for the dynamical regime $\textbf{X}(t)$, whose symmetry is described by a given subgroup $G_j$ of the parent group $G_0$, we must demand $\textbf{X}(t)$ to satisfy~\eqref{InvCond}. This equation means that $N$\nobreakdash-dimensional vector $\textbf{X}(t)$ is invariant under the action of all elements of the group $G_j$. It is essential, that~\eqref{InvCond} can substantially restrict the number ($m$) of nonzero terms in the decomposition~\eqref{ConfVectDecompose}. In the case $m=1$, only one term survives in~\eqref{ConfVectDecompose}, and then $\textbf{X}(t)$ turns out to be the Rosenberg nonlinear normal mode~\eqref{RosNNMdef} and, therefore, describes periodic vibrations.

If $m>1$, Eq.~\eqref{ConfVectDecompose} represents an $m$\nobreakdash-dimensional \textit{bush} of NNMs. Quasi-periodic motion with $m$ basic frequencies in the Fourier spectrum corresponds to such dynamical object. The vibrational modes entering into the given bush interact with each other, and the time-evolution of their amplitudes are described by $m$ functions $c_1(t),\ldots,c_m(t)$. For their explicit determination, we have to solve $m$ nonlinear differential equations. Thus, the search for an exact solution for $N$\nobreakdash-dimensional Hamiltonian system is reduced to the same problem for the $m$\nobreakdash-dimensional Hamiltonian system where, in practice, $m \ll N$.

In the present paper, we consider only one-dimensional bushes (i.e.\ $m=1$), or simply the Rosenberg NNMs. Since these modes are found with the aid of group-theoretical methods, we call them symmetry-determined NNMs. Hereafter the words ``symmetry determined'' are omitted for brevity.

\section{Nonlinear normal modes for diamond}

The space symmetry group of diamond structure in the equilibrium state is $Fd\bar3m$ in the international notion, or $O_h^7$ in the Schoenflies notation. There are two carbon atoms in the primitive cell determined by the three lattice vectors $\textbf{a}_1=(a,a,0), \textbf{a}_2=(a,0,a), \textbf{a}_3=(0,a,a)$. Here we give the Cartesian coordinates of these vectors and $a=1.79$~\AA. Since face-centered Bravais lattice corresponds to the group $Fd\bar3m$, these vectors are directed from the corner of the unit Bravais cell to the centers of three adjacent faces intersecting in this corner.

There are two carbon atoms in the primitive cell, and therefore 8 atoms in the Bravais cell. These atoms are located at the sites of lattice and at the four additional points inside the cell whose Cartesian coordinates are $(\frac14,\frac14,\frac14)$, $(\frac34,\frac34,\frac14)$, $(\frac34,\frac14,\frac34)$, $(\frac14,\frac34,\frac34)$ for the cubic cell with the unit edges.

When carbon atoms begin to oscillate near equilibrium positions of the diamond structure, the symmetry ($G_j$) of the vibrational state becomes lower than the symmetry ($G_0=Fd\bar3m$) of that crystal in the equilibrium state. Below, we discuss in detail the vibrational regimes corresponding to two NNMs associated with the irreducible representations $11{-}7$ and $10{-}3$ of the diamond space group $Fd\bar3m$. Using our group-theoretical methods~\cite{Chechin-Alone}, we have found that the space groups $G_j$ of these modes are $R\bar3m$ and $P4_132$ (in the Schoenflies notation $D_{3d}^5$ and $O^7$, respectively) which are subgroups of the group $Fd\bar3m$.

\subsection{Irreducible representations of space groups}

The matrix representation of a given group is the group of square matrices put into a homomorphic correspondence to a given group. Each group has infinitely many representations of different dimensions, but all of them can be constructed in the form of direct sums of some ``basic blocks'', which are irreducible representations (irreps) of this group. The latter representations possess many very interesting and important properties, which are considered in textbooks on symmetry groups (see, for example,~\cite{ElliottDawber}).

Point groups, which are used for describing molecules, have finite numbers of irreducible representations. The space groups, used for the describing ideal crystals, have an infinite number of irreps, since these representations are characterized by two indices, one of which is a wave vector $\textbf{k}$ that varies continuously within the primitive cell of the reciprocal lattice (Brillouin zone is often chosen as such cell). The second (discrete) index is the number of the irrep of the symmetry group of the above wave vector $\textbf{k}$. The irreducible representations of the latter group are often called ``small irreps'', in contrast to the full irreps, which are determined by the \textit{star} of the vector $\textbf{k}$ (it is obtained by acting on this vector by all point elements of the considered space group). The irreducible representations of all $230$ space groups can be found (in a complexly coded form) in the handbook by \cite{Kovalev}.

To find all possible NNMs and their bushes in the considered crystal, we need only those irreps of its space group that correspond to the \textit{points of high symmetry} in the Brillouin zone \cite{Kovalev}. They correspond to vectors $\textbf{k}$, which do not contain arbitrary parameters. In the case of the diamond group $G_0=Fd\bar3m$, there are four such vectors (numeration according to the Kovalev handbook):
\begin{equation}
\textbf{k}_8 =\frac14\textbf{b}_1-\frac14\textbf{b}_2+\frac12\textbf{b}_3, \quad \textbf{k}_9 =\frac12(\textbf{b}_1+\textbf{b}_2+\textbf{b}_3),  \quad \textbf{k}_{10} = \frac12(\textbf{b}_1+\textbf{b}_2),  \quad \textbf{k}_{11} =0,
\end{equation}
where $\textbf{b}_i$ are periods of the reciprocal lattice.

The requirement \eqref{InvCond}, demanding the vibrational regime to be invariant relative to a given subgroup $G_j$ of the parent group $G_0$, is a very strong condition for the case of a high symmetry group $G_j$. As a result of its fulfilment, many terms of the expansion~\eqref{ConfVectDecompose}, corresponding to different irreps of the group $G_0$, are zero since $c_j(t) = 0$. The concrete set of nonzero coefficients and some relations between them can be found by specific group-theoretical methods~\cite{Chechin-Alone} developed earlier for studying so-called ``complete condensate of the order parameters'' \cite{PSS, Acta} in the framework of the theory of structural phase transitions\footnote{Some group-theoretical computer software, created for this goal, we use in studying nonlinear dynamics of the systems with discrete symmetry.}.

Another reason for $c_j(t)$ to be zero is connected with the inability to construct the basis vectors of a given irrep on \textit{atomic displacements} for the crystal structure under consideration. In terms of representations, this means that the irrep is not contained in the decomposition of the crystal \textit{mechanical representation} into irreducible representations. This fact can be established using standard group-theoretical methods.

The main restriction on the number of nonzero terms in the decomposition~\eqref{ConfVectDecompose} is the list of \textit{invariant vectors} which correspond to different irreps for fixed symmetry group\footnote{This subgroup itself is singled out by one of the invariant vectors of the so-called ``critical irrep''.}~$G_j$. Actually, we must go through each of the irreps in turn and check whether any of its modes (the ``primary mode'') can involve into the vibrational process some modes belonging to any other irreps as the ``secondary modes''.

Further, we omit the technical details of the group-theoretical analysis and present only the final results for two NNMs in the diamond crystal lattice.

As was already mentioned, NNMs are associated with the individual full irreducible representations of the space group $Fd\bar3m$, which correspond to the wave vectors $\textbf{k}$ of high symmetry: $\textbf{k}_{11}=0$, $\textbf{k}_{10}=\frac12(\textbf{b}_1+\textbf{b}_2)$, $\textbf{k}_9=\frac12(\textbf{b}_1+\textbf{b}_2+\textbf{b}_3)$, and $\textbf{k}_8= \frac14\textbf{b}_1-\frac14\textbf{b}_2+\frac12\textbf{b}_3$.
Ten irreps correspond to the vector $\textbf{k}_{11}$: $4$ one-dimensional, $2$ two-dimensional, and $4$ three-dimensional. Four six-dimensional irreps correspond to the vector $\textbf{k}_{10}$. Six irreps correspond to the vector $\textbf{k}_9$: $4$ four-dimensional and $2$ eight-dimensional. Two twelve-dimensional irreps correspond to the vector $\textbf{k}_8$.
In the present paper, we consider only the irreps of the vectors $\textbf{k}_{10}$ and $\textbf{k}_{11}$ with a view to find any nonlinear normal modes belonging to these irreps.

Two carbon atoms of the diamond structure are located at the Wyckoff position $2a$ of the primitive cell. The Cartesian coordinates of $8$ carbon atoms in the cubic Bravais cell with unit edges, which is four times larger than the primitive cell, are $(0,0,0)$, $(\frac12,\frac12,0)$, $(\frac12,0,\frac12)$, $(0,\frac12,\frac12)$, $(\frac14,\frac14,\frac14)$, $(\frac34,\frac34,\frac14)$, $(\frac34,\frac14,\frac34)$, $(\frac14,\frac34,\frac34)$.

From the above-listed $22$ irreps of wave vectors $\textbf{k}_{10}$ and $\textbf{k}_{11}$ only the following irreps enter into the mechanical representation constructed on the atomic displacements of the carbon structure (they are called ``permissible irreps''): $11{-}7$, $11{-}10$, $10{-}1$, $10{-}3$, $10{-}4$. Hereafter, the first number in the irrep symbol refers to the wave vector, while the second number is associated with the small irrep (the representation of the wave-vector group).

Each NNM represents a certain \textit{one-parametric} dynamical regime since a certain pattern of atomic displacements corresponds to it, and all these displacements depend on \textit{only one} parameter. This parameter is the amplitude of the considered NNM. The following conclusion can be deduced from this property. Every vibrational NNM must belong to \textit{one} irrep (no modes of any other representations should be involved in the dynamical process), and this irrep must enter \textit{only one} time into the decomposition of the mechanical representation into irreducible components. Moreover, if the irrep is multidimensional, the NNM must correspond to one of its one-parametric invariant vectors. Such cases are quite rare. Indeed, only five NNMs can exist in the diamond structure for the wave vectors $\textbf{k}_{10}$ and $\textbf{k}_{11}$. In the present paper, we study two of these modes. They belong to the three-dimensional irreps $11{-}7$ (mode~1) and to six-dimensional irrep $10{-}3$ (mode~2), respectively. Let us consider these nonlinear normal modes in more detail.

\subsection{Vibrational pattern of NNM~1}

As was already mentioned, the size of the primitive cell (and the Bravais cell) in the equilibrium and vibrational states do not coincide with each other in the general case. The size and shape of the cell in the crystal vibrational state are determined by the star of the wave vector of the considered irreducible representation.

\begin{figure}[h]
    \begin{minipage}[h]{0.47\linewidth}
        \center{\includegraphics[width=1\linewidth]{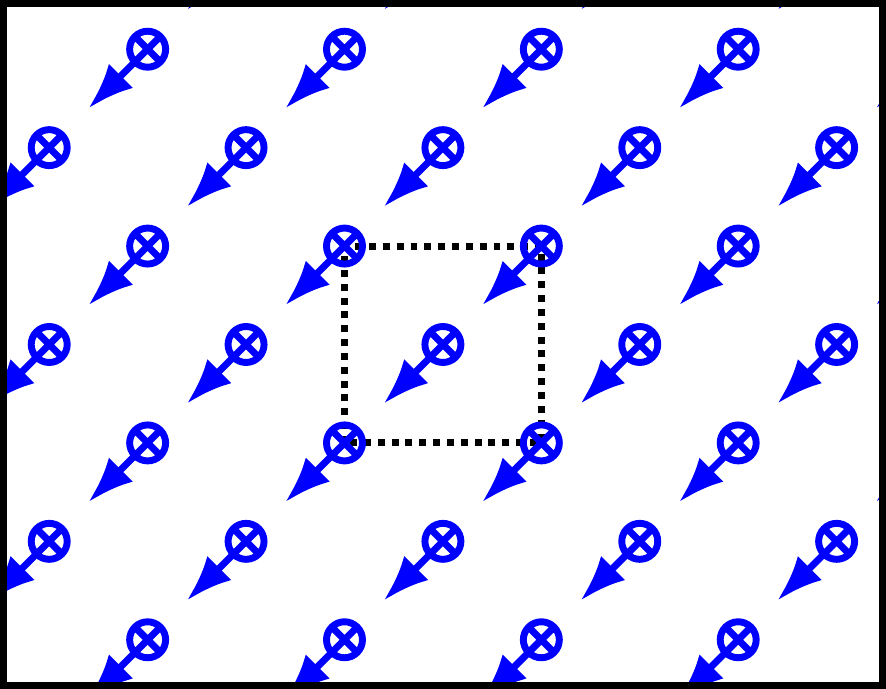} \\ (a) $z=0$}
    \end{minipage}
    \hfill
    \begin{minipage}[h]{0.47\linewidth}
        \center{\includegraphics[width=1\linewidth]{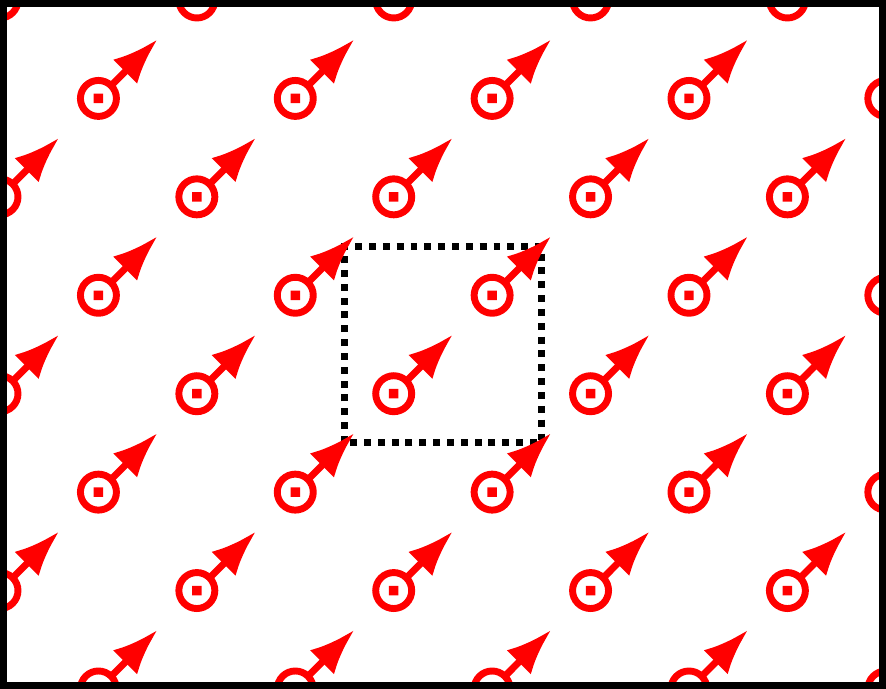} \\ (b) $z=\frac12a$}
    \end{minipage}
	\vspace{.75\baselineskip}\par
	\begin{minipage}[h]{0.47\linewidth}
		\center{\includegraphics[width=1\linewidth]{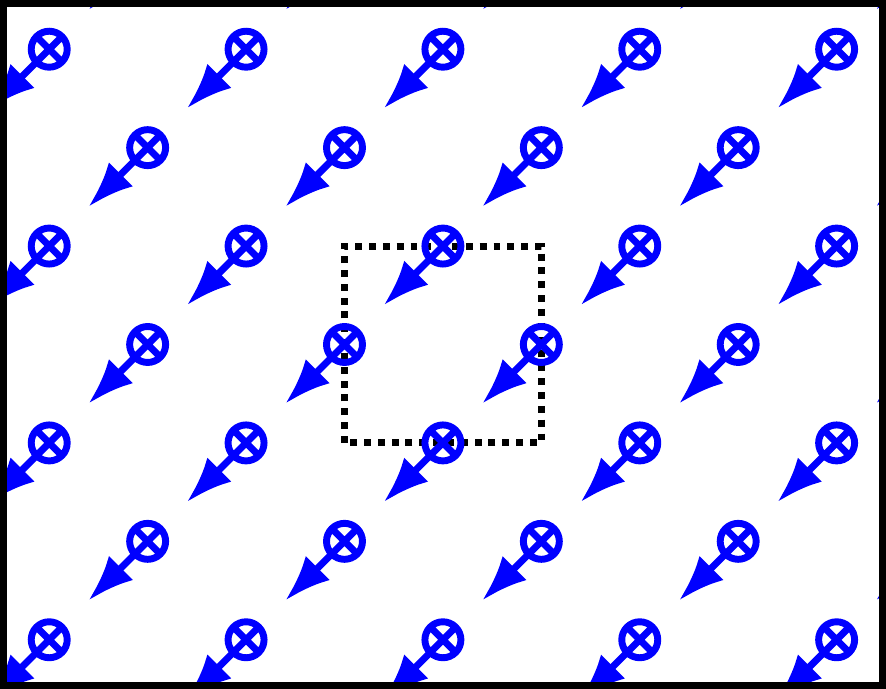}} \\ (c) $z=a$
	\end{minipage}
	\hfill
	\begin{minipage}[h]{0.47\linewidth}
		\center{\includegraphics[width=1\linewidth]{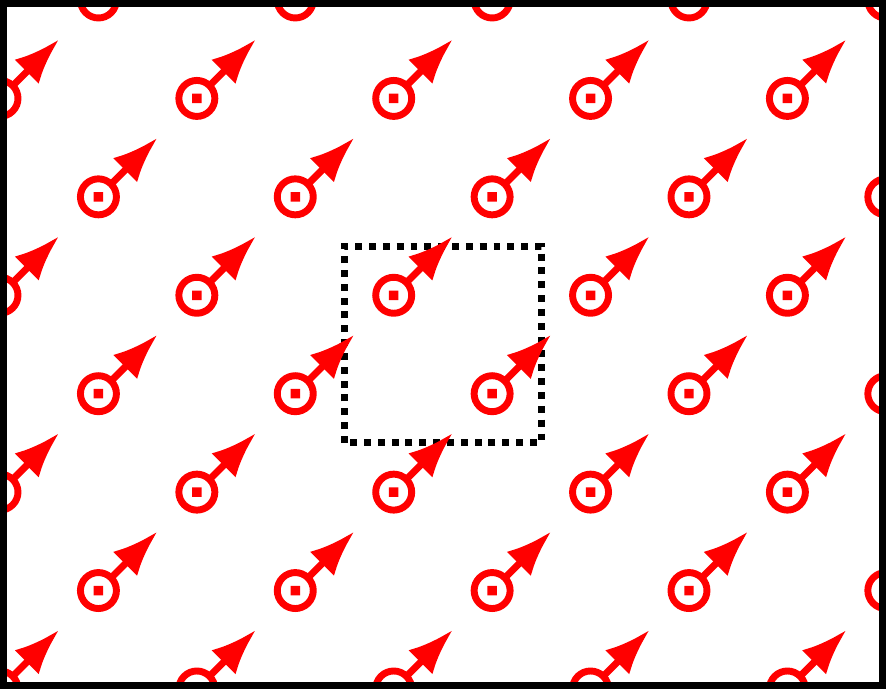}} \\ (d) $z=\frac32a$
	\end{minipage}
	\caption{3D vibrational pattern of mode~1 as a stack of flat grids perpendicular to the $Z$\nobreakdash-axis. Coordinates are given in units of the lattice constant $a=1.79$~\AA.}
	\label{fig2}
\end{figure}

The single-arm star of the vector $\textbf{k}_{11}=0$ corresponds to the irrep $11{-}7$. This means that the size of vibrational cell is equal to that of the diamond equilibrium cell. For clarity, we present the 3D vibrational patterns in Fig.~\ref{fig2} as a stack of flat grids perpendicular to the $Z$\nobreakdash-axis. The displacements of carbon atoms in $XY$\nobreakdash-plane are depicted by arrows, while $z$\nobreakdash-displacements are designated, as is customary in crystallography, by circles with a dot (direction along the $Z$\nobreakdash-axis) and circles with a cross (direction opposite to the $Z$\nobreakdash-axis). It is essential to note that values of all displacements along $X$\nobreakdash-, $Y$\nobreakdash-, and $Z$- directions are the \textit{same}.

The atomic displacements for atoms in one primitive cell are presented in Table~\ref{table1}.

All atomic displacements in Fig.~\ref{fig2} are depicted for a certain instant~$t_0$. In this figure, one can see several atomic layers (plane grids) located at different distances from each other along the axis~$Z$. Displacements of all atoms in every layer are \textit{identical}. Note that there is an alternation of $z$\nobreakdash-component directions of all atoms when we pass from one layer to the next.

The described character of displacements of atoms belonging to each given layer allows us to speak about displacement of the \textit{entire layer} as a single whole. The dynamical regime described by the considered NNM can be obtained by multiplying the given pattern of atomic displacements by some time-periodic function [see Eq.~\eqref{RosNNMdef}]. We discuss the properties of this function in the next Section. Thus, adjacent atomic layers oscillate in antiphase with respect to each other, and these oscillations occur simultaneously along $X$\nobreakdash-, $Y$\nobreakdash-, and $Z$\nobreakdash-axes.

In horizontal plane (see Fig.~\ref{fig2}), all atoms move along diagonal $XY$\nobreakdash-direction of the Bravais cell. However, one can consider the equivalent $X\overline{Y}$ direction\footnote{Here $X\overline{Y}$ means positive displacement along $X$\nobreakdash-axis and negative displacement along $Y$\nobreakdash-axis.} corresponding to another diagonal. There are no physical reasons for assigning special role to any of these directions (they are symmetrically equivalent). It means that there must exist a~\textit{twin} of the NNM with pattern shown in Fig.~\ref{fig2}, whose atoms move along $X\overline{Y}$ direction. All dynamical properties of NNM-twins are identical. In the theory of bushes of nonlinear normal modes, such modes (or their bushes) are called ``dynamical domains'' by analogy with the corresponding notion of the theory of structural phase transitions. These objects possess identical dynamical properties and transform into each other under action of the symmetry elements, which disappear when we pass from the parent symmetry group $G_0$ to its subgroup $G_j$. In general case, there exist $n$ dynamical domains, and this number is equal to the index of the subgroup $G_j$ in the group $G_0$, i.e.\ $n=\|G_0\|/\|G_j\|$ where $\|G_0\|$ and $\|G_j\|$ are orders of the corresponding groups. Note that in the present case $G_0=Fd\bar3m$, $G_j=R\bar3m$, and $n=4$.

\begin{figure}[h]
	\begin{minipage}[h]{0.47\linewidth}
		\center{\includegraphics[width=1\linewidth]{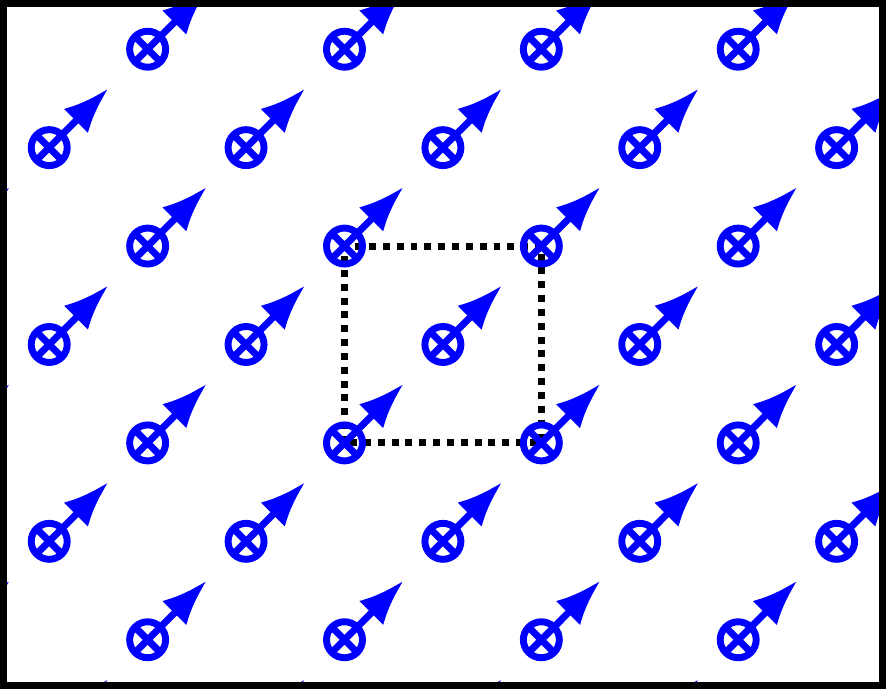} \\ (a) $z=0$}
	\end{minipage}
	\hfill
	\begin{minipage}[h]{0.47\linewidth}
		\center{\includegraphics[width=1\linewidth]{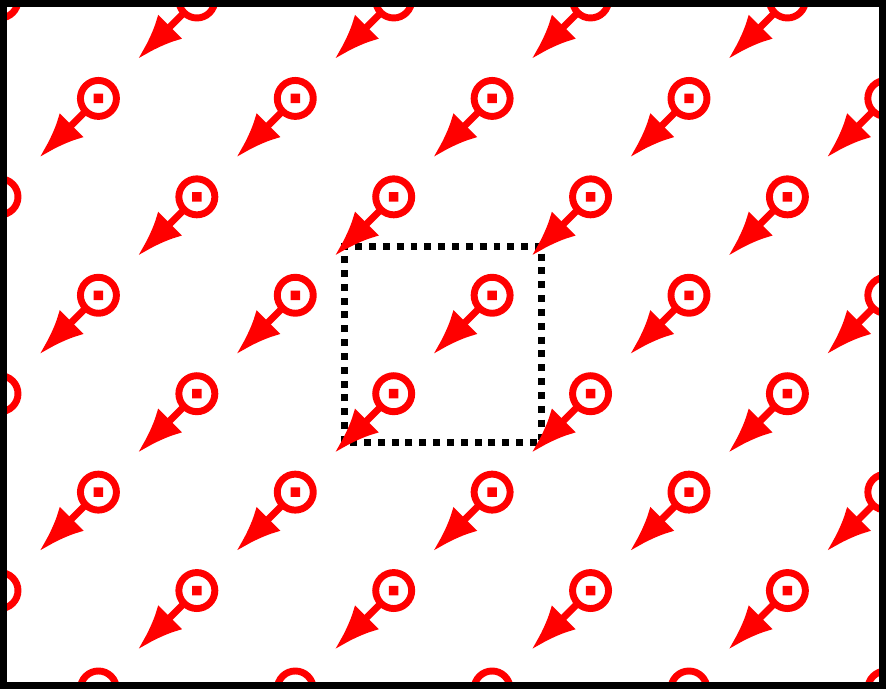} \\ (b) $z=\frac12a$}
	\end{minipage}
	\vspace{.75\baselineskip}\par
	\begin{minipage}[h]{0.47\linewidth}
		\center{\includegraphics[width=1\linewidth]{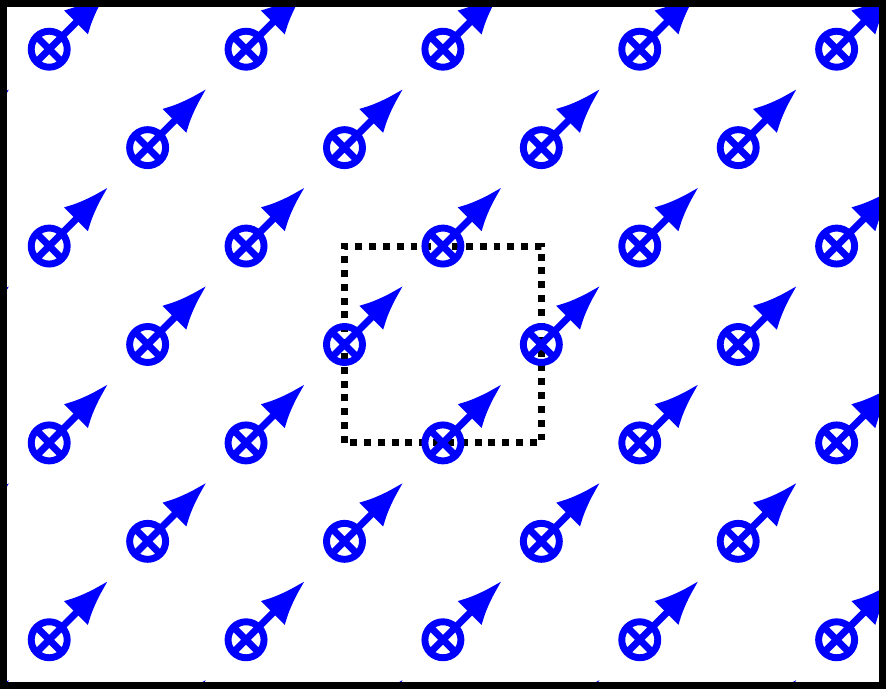}} \\ (c) $z=a$
	\end{minipage}
	\hfill
	\begin{minipage}[h]{0.47\linewidth}
		\center{\includegraphics[width=1\linewidth]{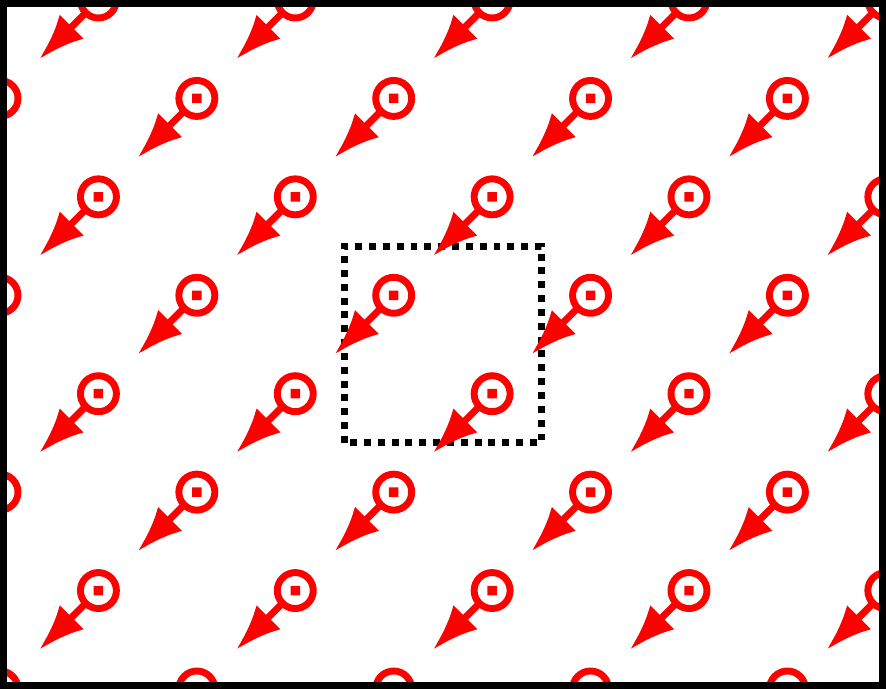}} \\ (d) $z=\frac32a$
	\end{minipage}
	\caption{One of vibrational domains of mode~1 as a stack of flat grids perpendicular to the $Z$\nobreakdash-axis. Coordinates are given in units of the lattice constant $a=1.79$~\AA.}
	\label{fig2a}
\end{figure}

As a consequence, the symmetry of the diamond vibrational state becomes four times lower than that of equilibrium state because rotations by $180^{\circ}$ angles about $X$\nobreakdash-, $Y$\nobreakdash-, and $Z$\nobreakdash-axes disappear when we pass from the parent group $G_0$ to its subgroup $G_j$. Therefore, three dynamical domains additional to the domain shown in Fig.~\ref{fig2} must exist in diamond structure. In Fig.~\ref{fig2a} we present one of these domains, which can be obtained by action on the pattern in Fig.~\ref{fig2} by $180^{\circ}$-rotation about $Z$\nobreakdash-axis ($z$\nobreakdash-component of atomic displacements does not change, while $x$- and $y$\nobreakdash-components change their signs).

\subsection{Vibrational pattern of NNM~2}

Nonlinear normal mode~2 belongs to irrep $10{-}3$ corresponding to the star of wave vector~$\textbf{k}_{10}$. The arms of this star are vectors $(\frac12\textbf{b}_1+\frac12\textbf{b}_2)$, $(\frac12\textbf{b}_1+\frac12\textbf{b}_3)$, and $(\frac12\textbf{b}_2+\frac12\textbf{b}_3)$, which can be obtained by action on the vector $\textbf{k}_{10}$ of all elements of the parent group $Fd\bar3m$. Such form of the star dictates \textit{doubling} of the edges $\textbf{a}_1$, $\textbf{a}_2$, $\textbf{a}_3$ of the primitive cell, and, consequently, doubling of the Bravais-cell edges when we pass from equilibrium to vibrational state of the diamond structure. But actually, this NNM preserves translation vectors of original Bravais cell [$\textbf{a}'_1=-\textbf{a}_1+\textbf{a}_2+\textbf{a}_3=(2a,0,0)$, etc.] which may be considered as lattice vectors of the \textit{extended cell} corresponding to the mode~1.

\begin{figure}[h]
	\begin{minipage}[h]{0.47\linewidth}
		\center{\includegraphics[width=1\linewidth]{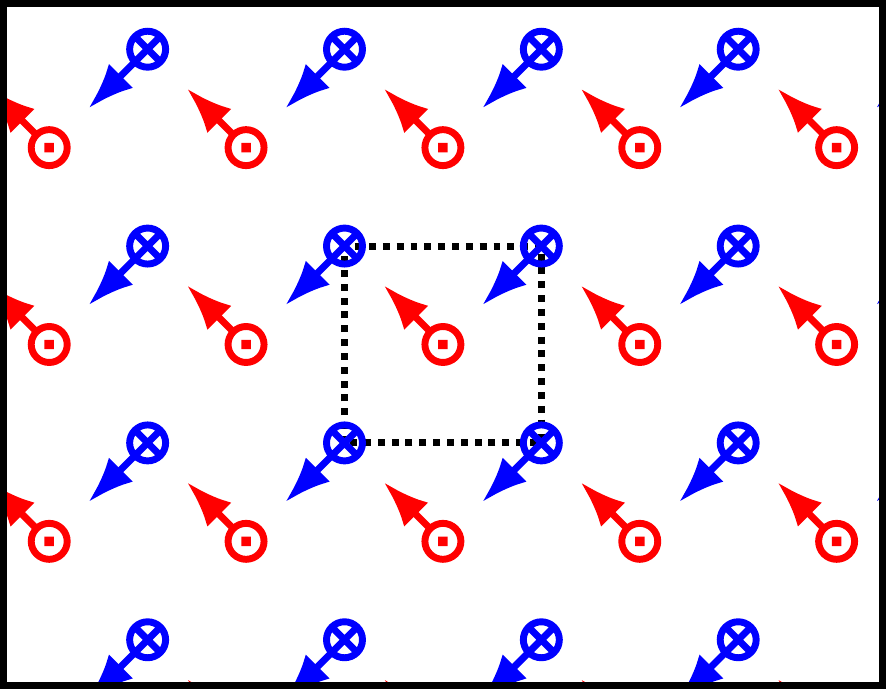}} \\ (a) $z=0$
	\end{minipage}
	\hfill
	\begin{minipage}[h]{0.47\linewidth}
		\center{\includegraphics[width=1\linewidth]{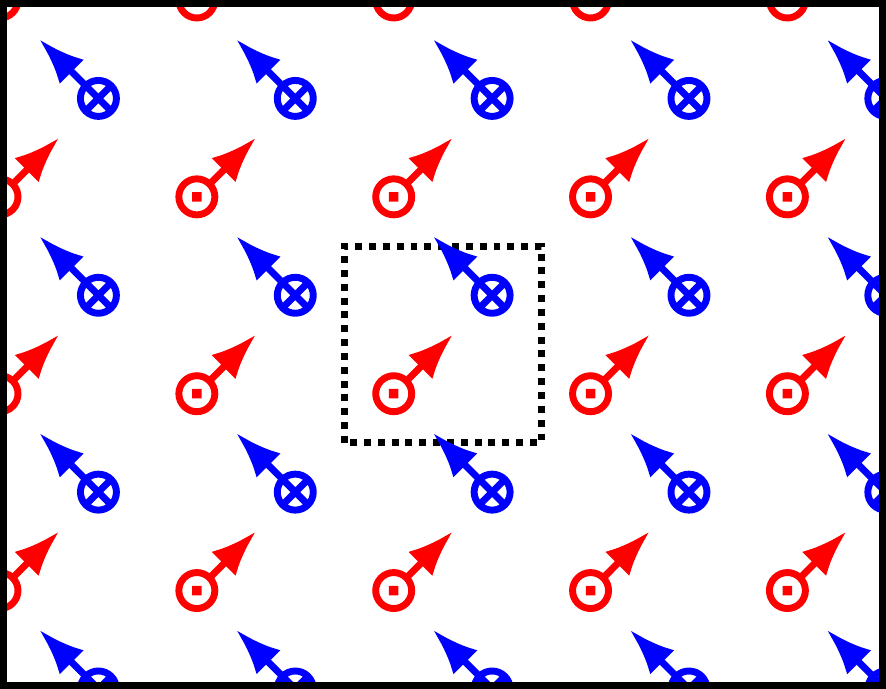}} \\ (b) $z=\frac12a$
	\end{minipage}
	\vspace{.75\baselineskip}\par
	\begin{minipage}[h]{0.47\linewidth}
		\center{\includegraphics[width=1\linewidth]{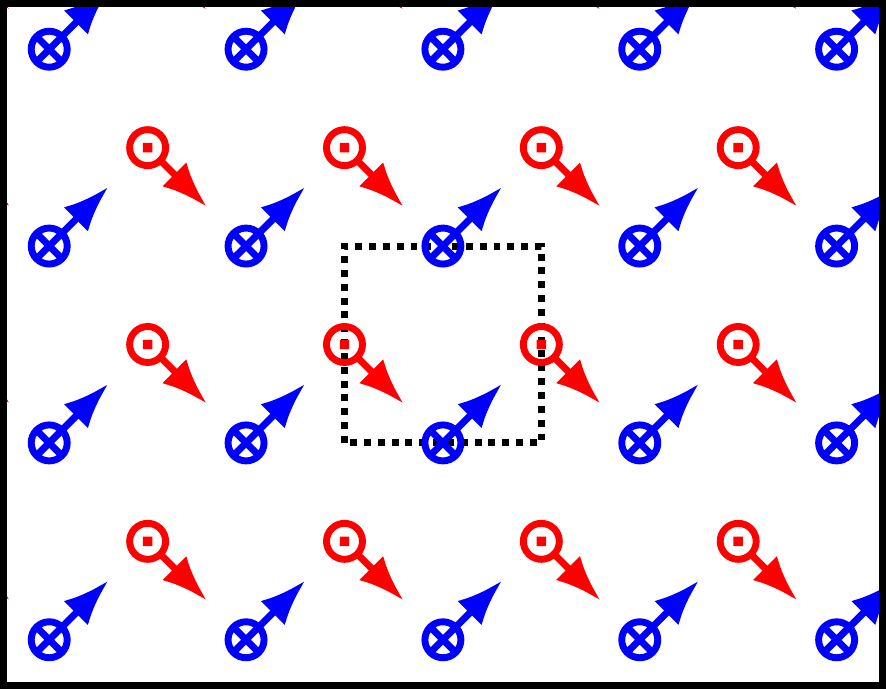}} \\ (c) $z=a$
	\end{minipage}
	\hfill
	\begin{minipage}[h]{0.47\linewidth}
		\center{\includegraphics[width=1\linewidth]{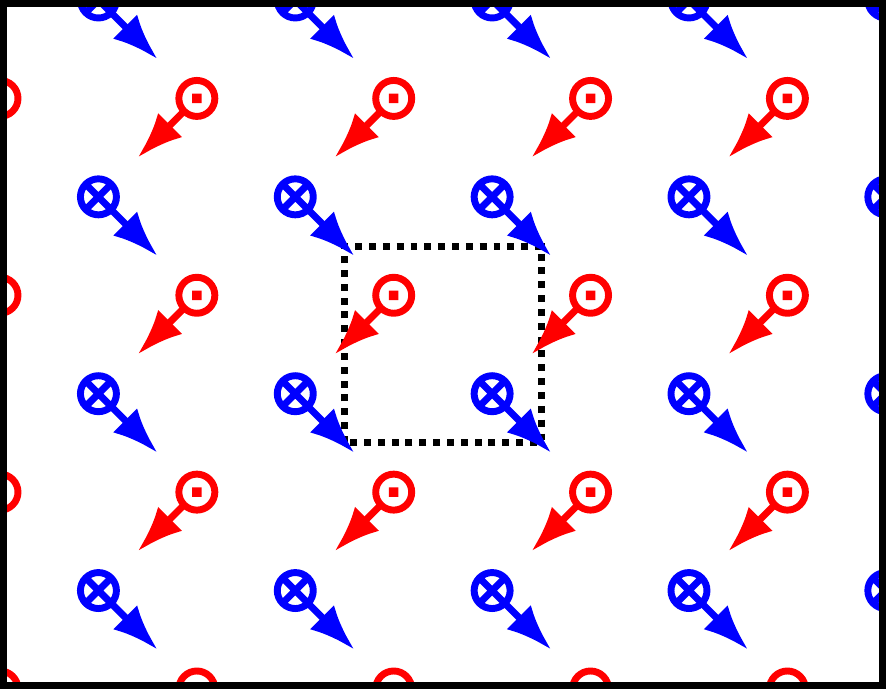}} \\ (d) $z=\frac32a$
	\end{minipage}
	\caption{3D vibrational pattern of mode~2 as a stack of flat grids perpendicular to the $Z$\nobreakdash-axis. Coordinates are given in units of the lattice constant $a=1.79$~\AA.}
	\label{fig3}
\end{figure}

Let us consider the two-dimensional patterns of atomic displacements presented in Fig.~\ref{fig3}. As in Fig.~\ref{fig2} of the patterns of NNM~1, we use here the same notation for atomic displacements. Absolute values of these displacements along $X$\nobreakdash-, $Y$\nobreakdash-, and $Z$\nobreakdash-axes are equal to each other.

All atoms in the plane patterns of mode~1 (see Fig.~\ref{fig2}) are displaced along $XY$ diagonal direction. Unlike this picture, the plane patterns in  Fig.~\ref{fig3} for mode~2 demonstrate atomic displacements along both $XY$ and $X\overline{Y}$ diagonal directions. Moreover, along each of these directions, there is alternation of $XY$ and $X\overline{Y}$ displacements, as well as alternation of the sign of $z$-displacements. As the result, the distance between neighbouring atoms with equivalent displacements in each of the above directions is twice larger than that in the diamond
equilibrium state. The atomic displacements of all atoms in one Bravais cell are presented in Table~\ref{table2}.

The time evolution of atomic displacements of the three-dimensional pattern of mode~2 is discussed in the next section of the paper.

\section{Dynamics of diamond lattice corresponding to nonlinear normal modes}

Choosing an arbitrary pattern of atomic displacements as initial conditions for solving nonlinear dynamical equations of the considered Hamiltonian system, we will obtain very quickly that this pattern doesn't survive during time-evolution. In contrast, we should expect that the pattern corresponding to any bush of modes, in particular to nonlinear normal mode, will persist infinitely long, since any bush is an exact solution of these equations.

The verification of these conclusions of the theory of bushes of nonlinear normal modes can be carried out by the methods of molecular dynamics for those Hamiltonian systems for which sufficiently reliable potentials of interparticle interactions are known. Unfortunately, such potentials are usually either unknown or turn out to be very complex because they are determined by huge tables of phenomenological parameters entering into these potentials (see, for example, \cite{Brenner}).

A more adequate way to verify group-theoretical results is to apply methods of the density functional theory (DFT)~\cite{Kohn}. We already used this approach for studying large-amplitude atomic vibrations in graphene~\cite{graphene-2}, graphane~\cite{graphane}, carbyne~\cite{carbyne}, and molecule $SF_6$~\cite{SF6}. In the present paper, we use DFT-methods to study dynamics of the above discussed NNMs in diamond. For this purpose, we have used the software package Quantum Espresso~\cite{QE}. Below, some results of studying NNMs 1 and 2 are discussed.

\begin{figure}[h]
	\begin{minipage}[h]{0.49\linewidth}
		\center{\includegraphics[width=1\linewidth]{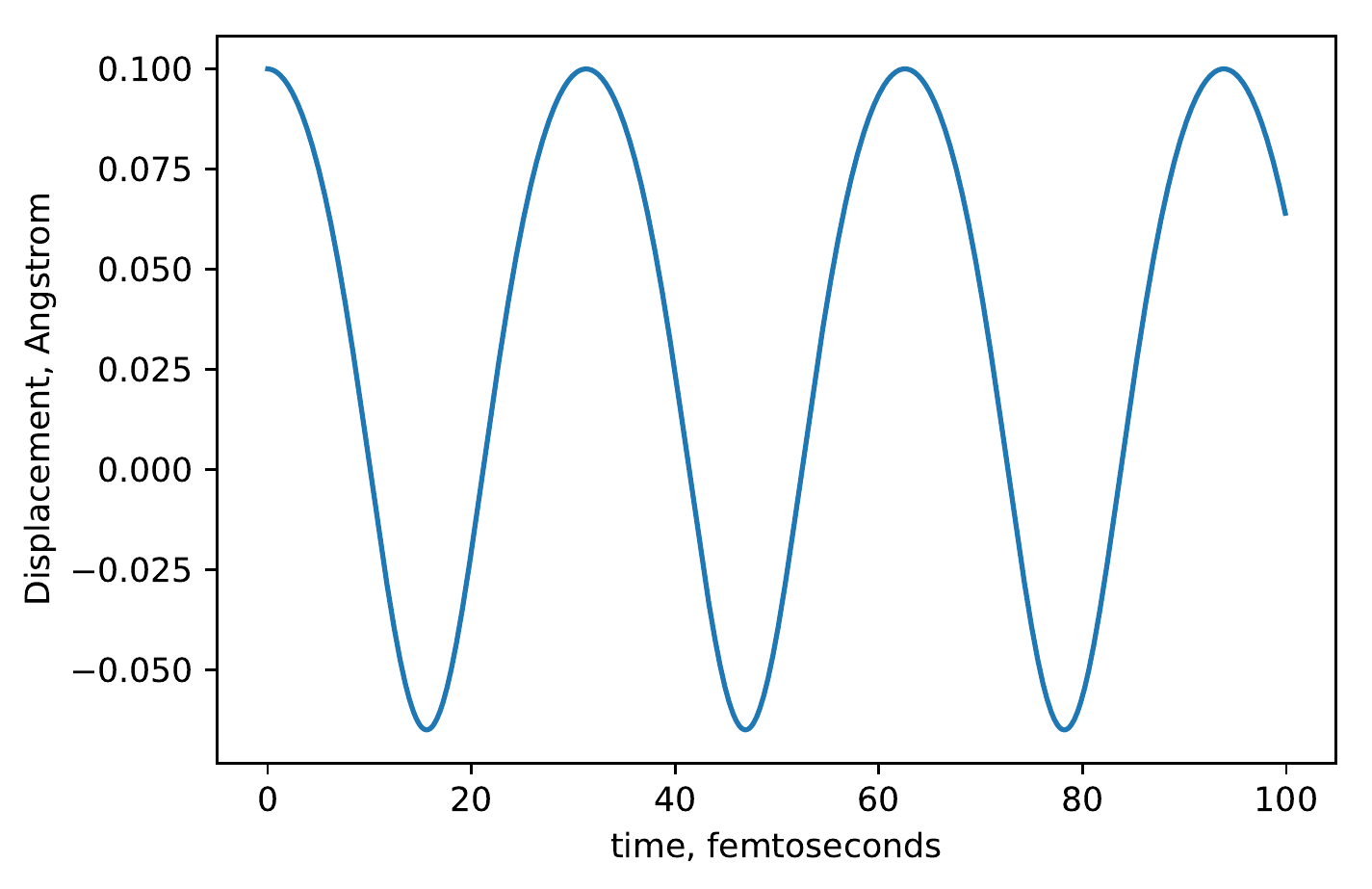} \\ (a)}
	\end{minipage}
	\hfill
	\begin{minipage}[h]{0.49\linewidth}
		\center{\includegraphics[width=1\linewidth]{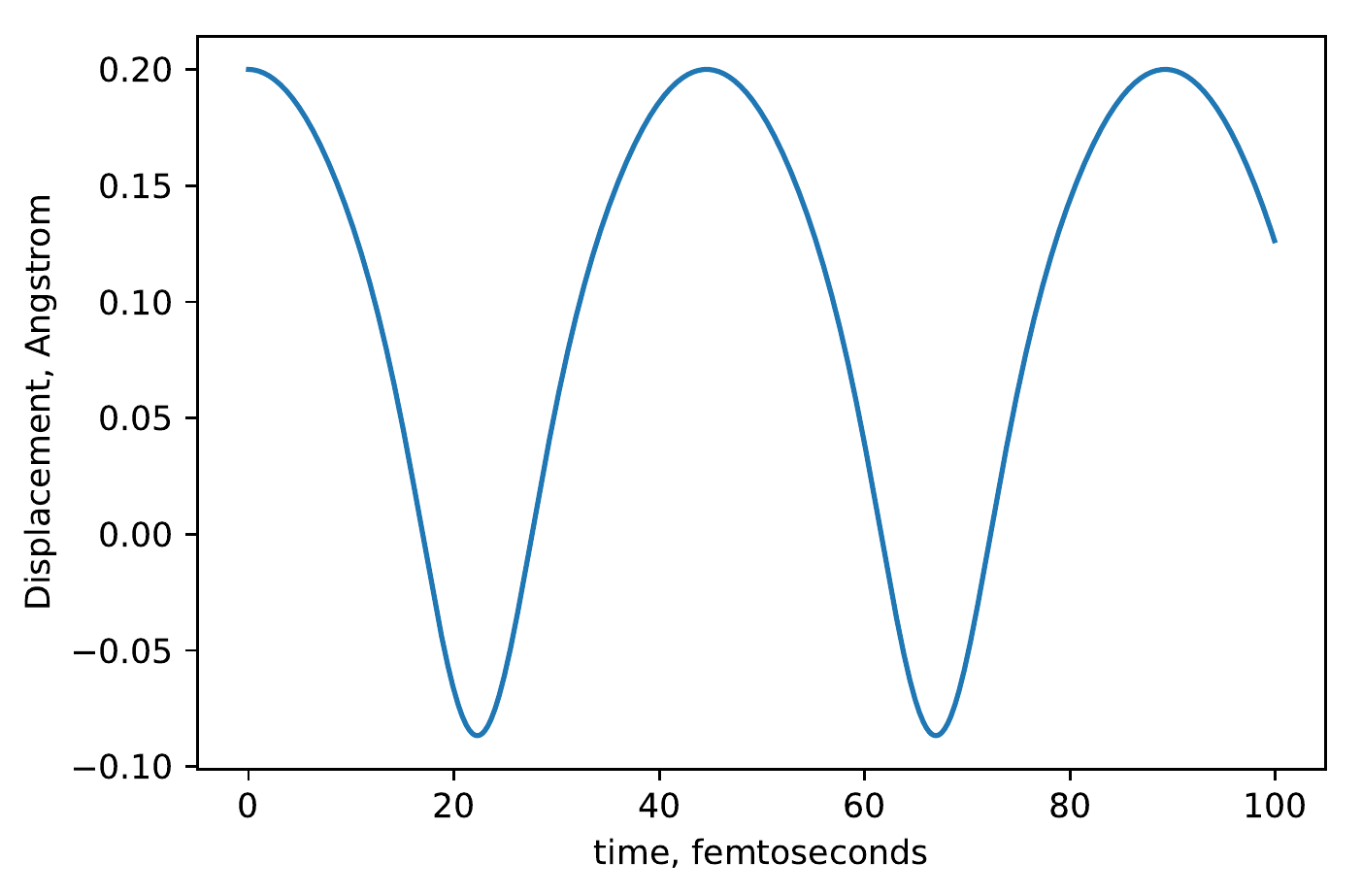} \\ (b)}
	\end{minipage}
	\caption{Atomic oscillations, corresponding to NNM~1 for two amplitude values: (a)~0.1~\AA\ and (b)~0.2~\AA.}
	\label{fig4}
\end{figure}

In Fig.~\ref{fig4}, vibrations of one of the carbon atoms of the diamond structure for NNM~1 are presented for two values of its amplitude. Note that time-dependence for all other atoms of this crystal is the same due to the definition \eqref{RosNNMdef} of NNM.
The nonlinearity of interatomic interactions in diamond must affect the considered vibrational modes in two ways. Indeed, we know from different methods of nonlinear dynamics that for the case of small nonlinearity one has to search power series with respect to small parameter not only for the \textit{form} of oscillations, but also for their \textit{frequency} to avoid the secular terms.

\begin{figure}[h]
	\centerline{\includegraphics[width=0.5\linewidth]{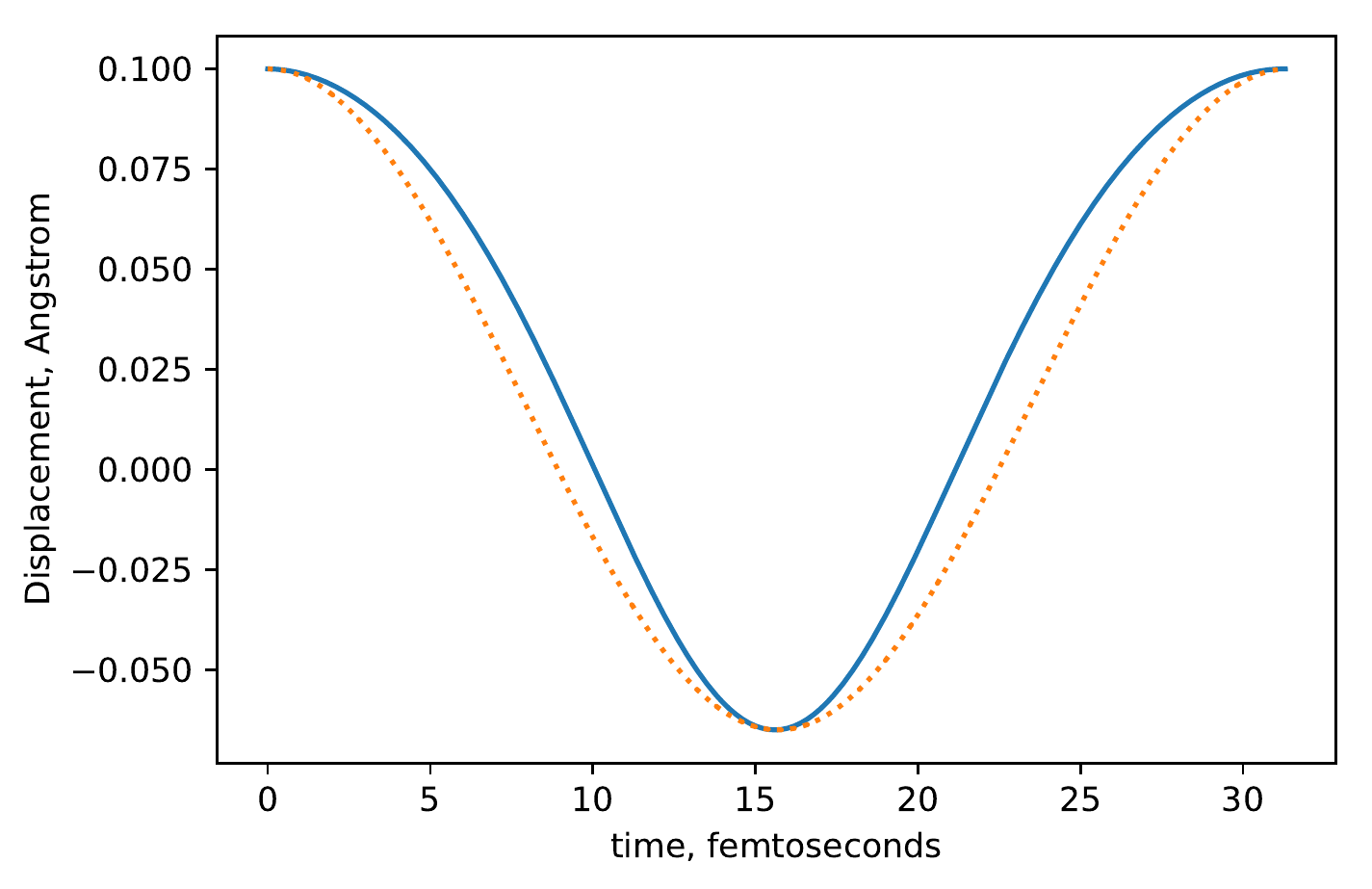}}
	\vspace*{3pt}
	\caption{Fitting of the oscillations from Fig.~\ref{fig4}a, corresponding to NNM~1 (solid line) by sinusoidal function (dotted line).}
	\label{fig5}
\end{figure}

\begin{figure}[h]
	\begin{minipage}[h]{0.49\linewidth}
		\center{\includegraphics[width=1\linewidth]{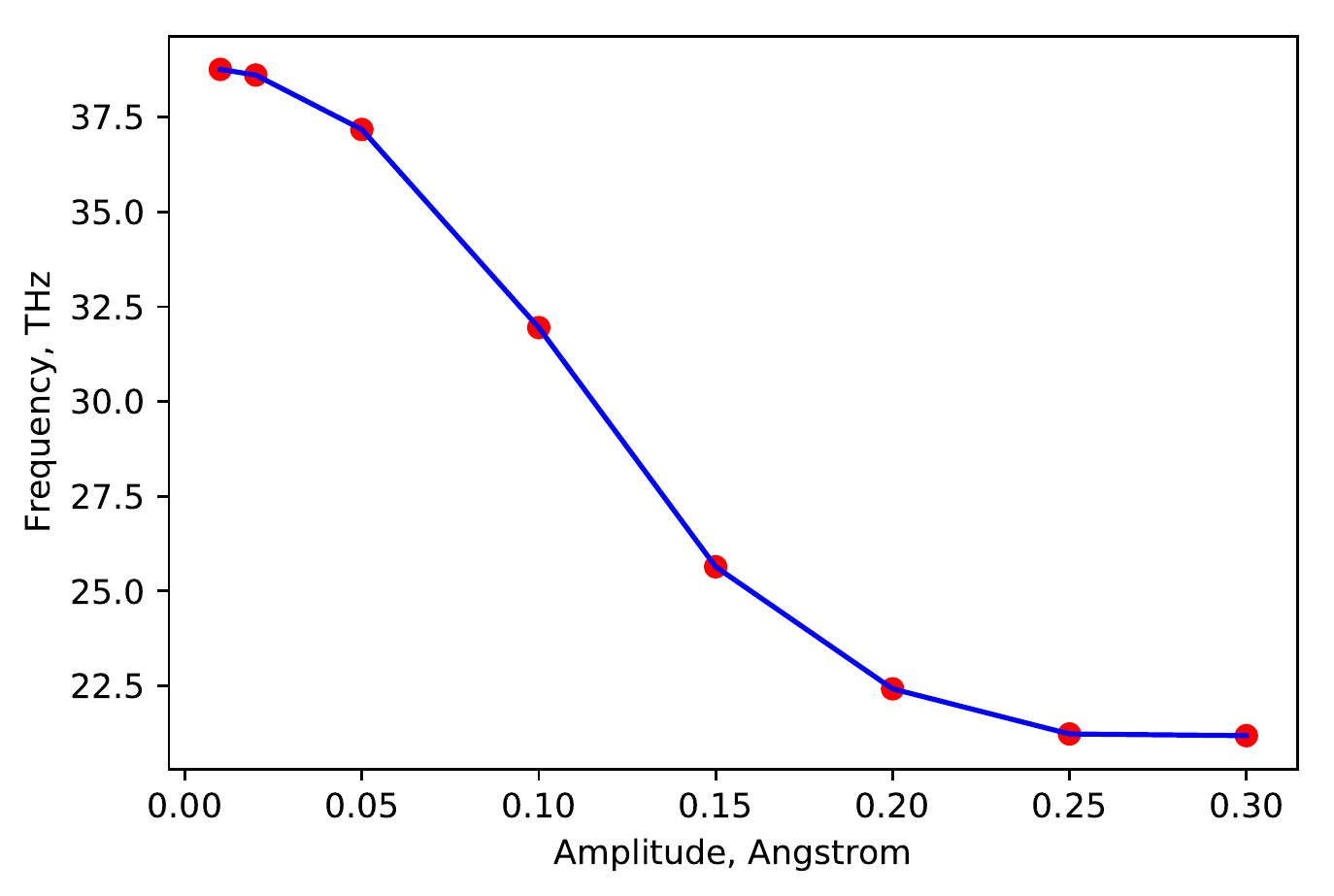} \\ (a)}
	\end{minipage}
	\hfill
	\begin{minipage}[h]{0.49\linewidth}
		\center{\includegraphics[width=1\linewidth]{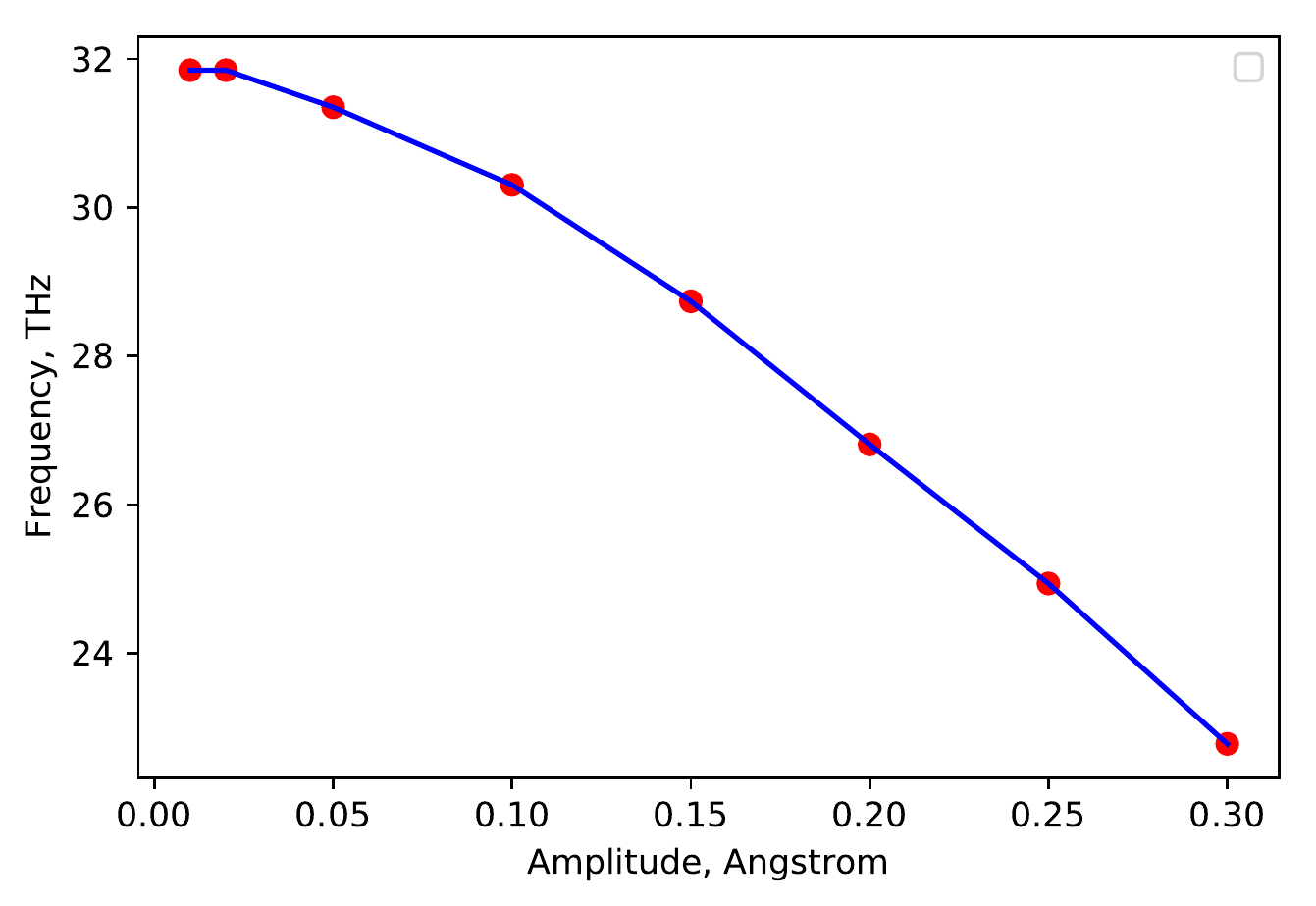} \\ (b)}
	\end{minipage}
	\caption{Dependence of frequency on amplitude for (a)~mode~1 and (b)~mode~2.}
	\label{fig6}
\end{figure}

In Fig.~\ref{fig5}, we show a fitting to the plot in Fig.~\ref{fig4}a (over one period) by sinusoidal function. It is obvious from this figure, that the anharmonism of the diamond crystal exerts \textit{small} influence on the form of atomic vibrations associated with mode~1. However, the anharmonism shows itself very \textit{strong} in the dependence of frequency on amplitude, which can be seen in Fig.~\ref{fig6}a. Indeed, this dependence demonstrates a \textit{soft} type of nonlinearity and the frequency of mode~1 decreases from $37.17$~THz to $21.19$~THz when the amplitude increases from $0.05$~\AA\ to $0.3$~\AA\ (note that in the case of pure harmonic oscillations there is no dependence of the frequency on amplitude).

\begin{figure}[h]
	\begin{minipage}[h]{0.32\linewidth}
		\center{\includegraphics[width=1\linewidth]{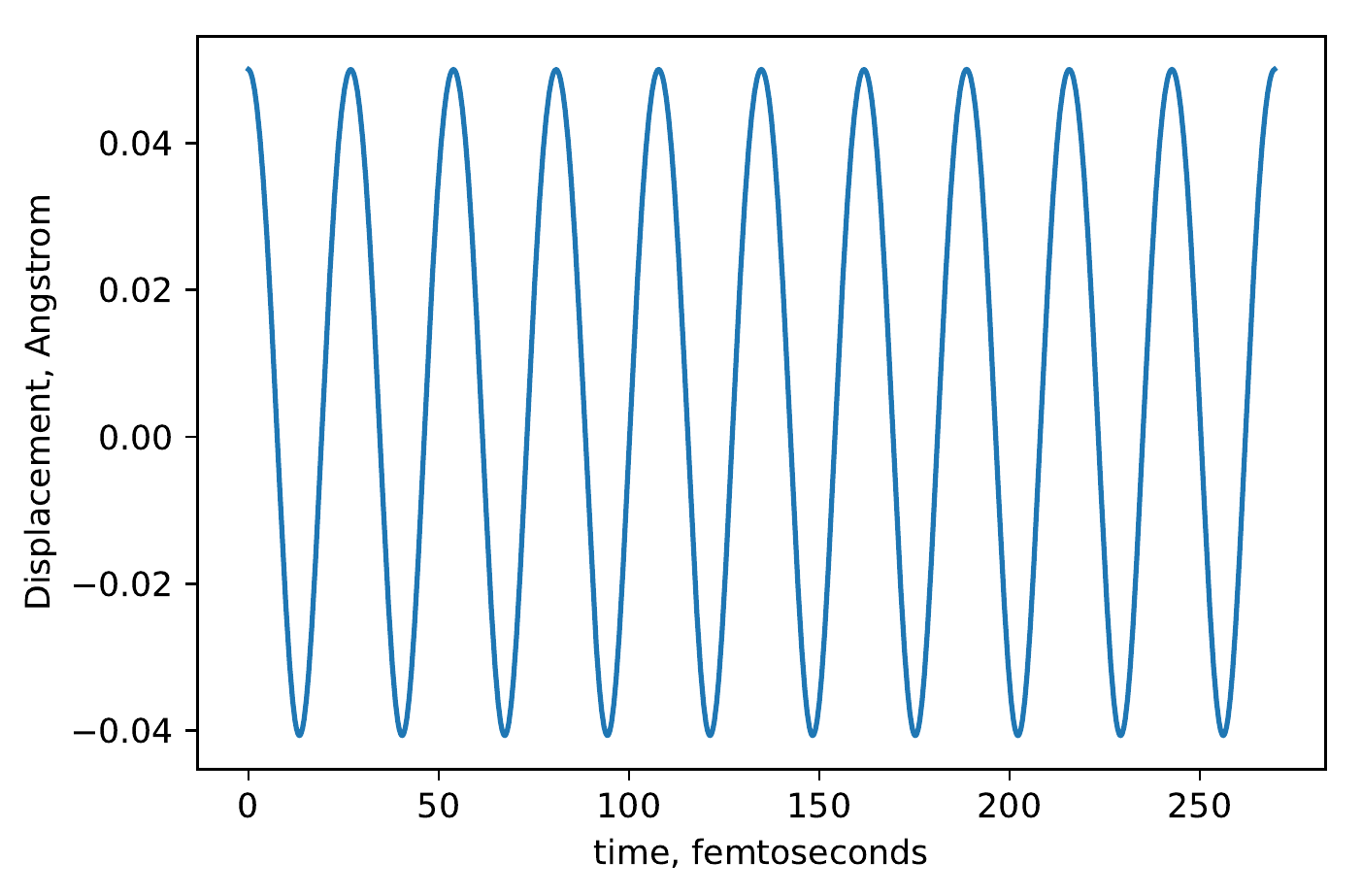} \\ (a)}
	\end{minipage}
	\hfill
	\begin{minipage}[h]{0.32\linewidth}
		\center{\includegraphics[width=1\linewidth]{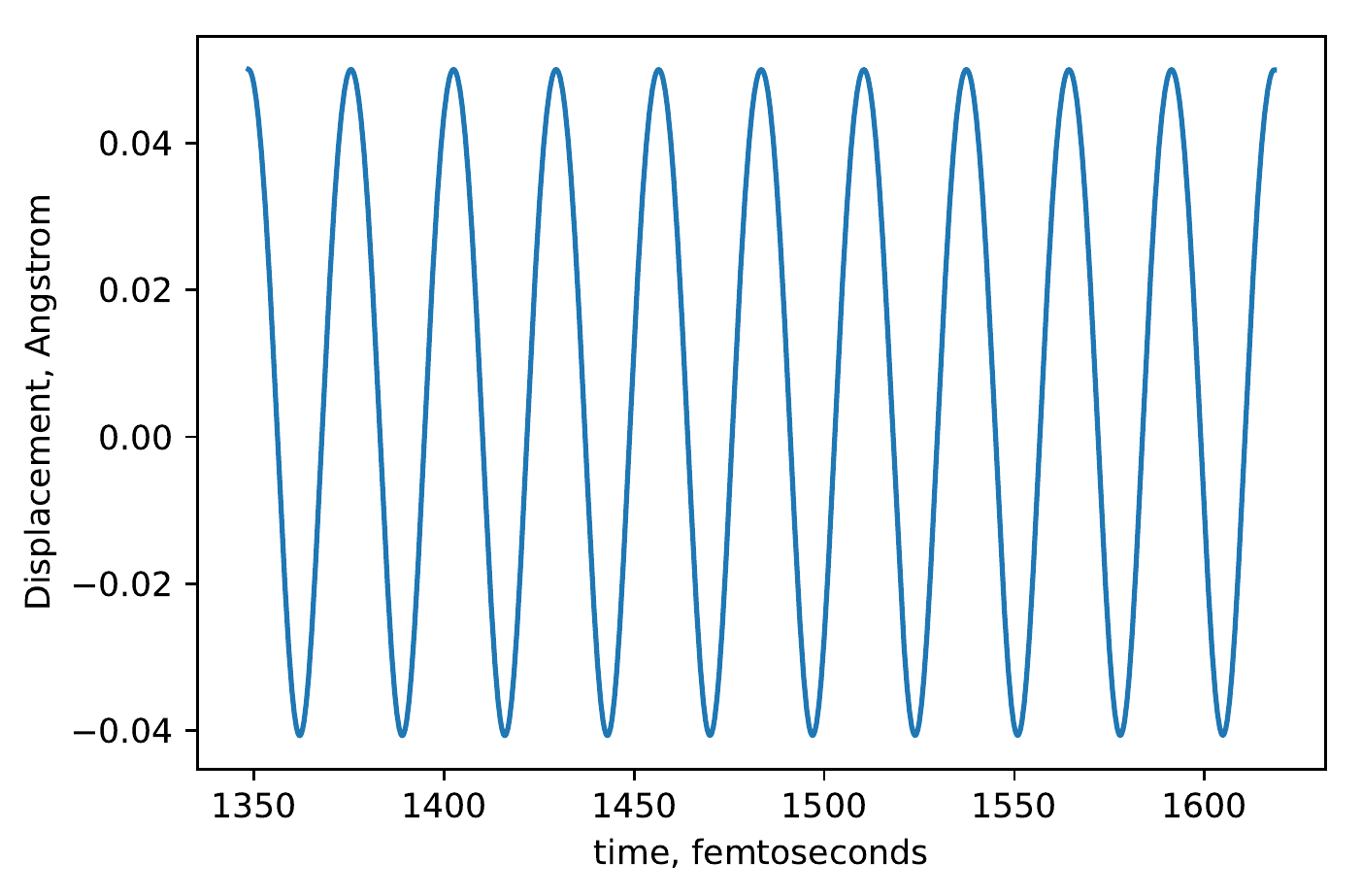} \\ (b)}
	\end{minipage}
	\hfill
	\begin{minipage}[h]{0.32\linewidth}
		\center{\includegraphics[width=1\linewidth]{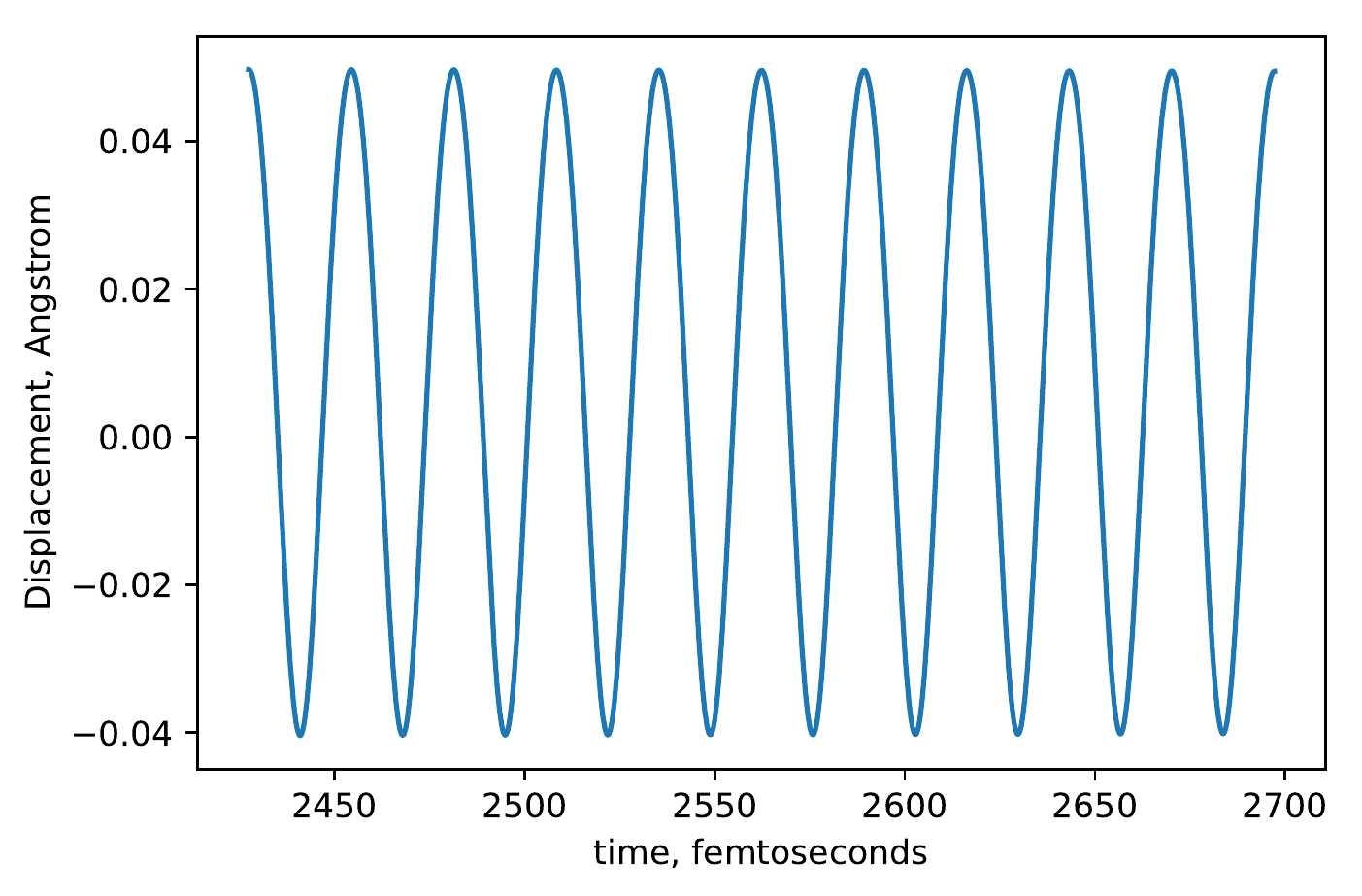} \\ (c)}
	\end{minipage}
	\caption{Atomic oscillations, corresponding to mode~1 with amplitude $0.05$~\AA\ for three different subintervals: (a)~$0\isep 10T$; (b)~$50T\isep 60T$; (c)~$90T\isep 100T$, where $T$ is the period of oscillations.}
	\label{fig7}
\end{figure}

\begin{figure}[h]
	\centerline{\includegraphics[width=0.5\linewidth]{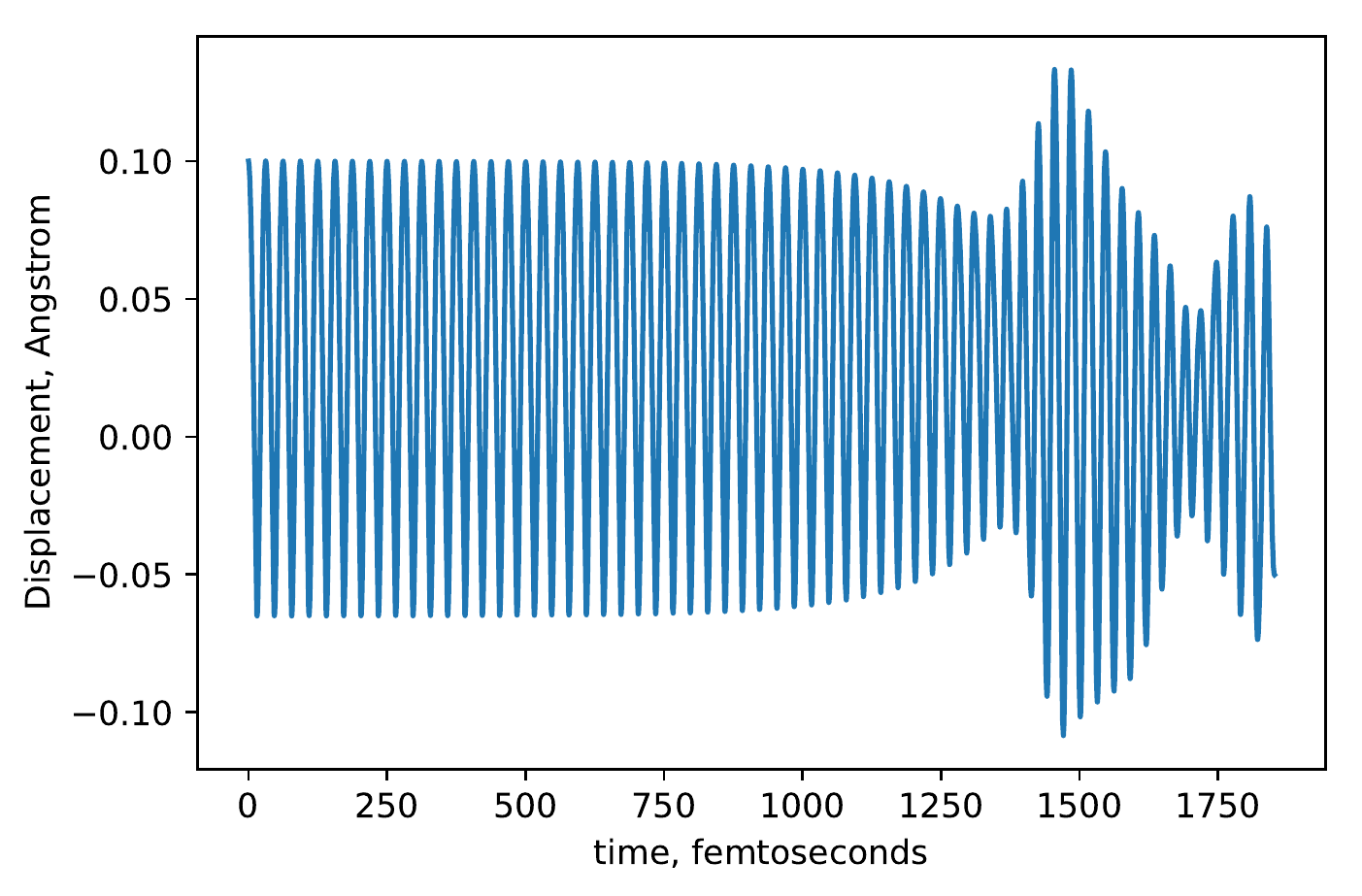}}
	\vspace*{3pt}
	\caption{Atomic oscillations, corresponding to mode~1 with amplitude $0.2$~\AA.}
	\label{fig8}
\end{figure}

Let us note that the frequencies of both vibrational modes, discussed in the present paper, are hit into the diamond phonon spectrum. Despite this fact, the mode~1 occurs to be a sufficiently stable dynamical object for the amplitudes up to the value $0.05$~\AA. Indeed, its form does not change essentially within time intervals corresponding to $100$ periods of oscillations (see Fig.~\ref{fig7}). However, for larger amplitudes it loses its stability. This fact can be seen in Fig.~\ref{fig8}.

\begin{figure}[h]
	\begin{minipage}[h]{0.32\linewidth}
		\center{\includegraphics[width=1\linewidth]{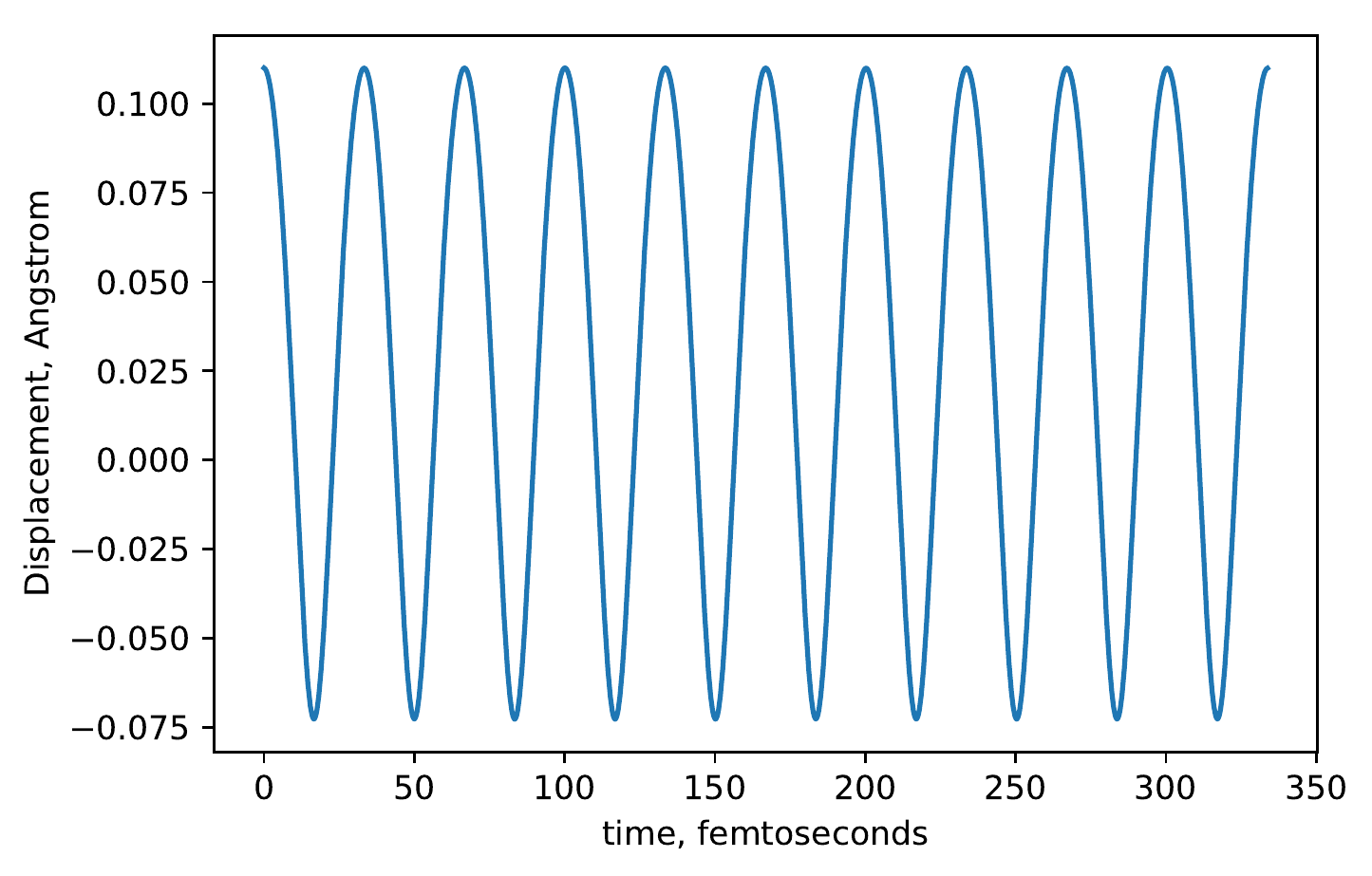} \\ (a)}
	\end{minipage}
	\hfill
	\begin{minipage}[h]{0.32\linewidth}
		\center{\includegraphics[width=1\linewidth]{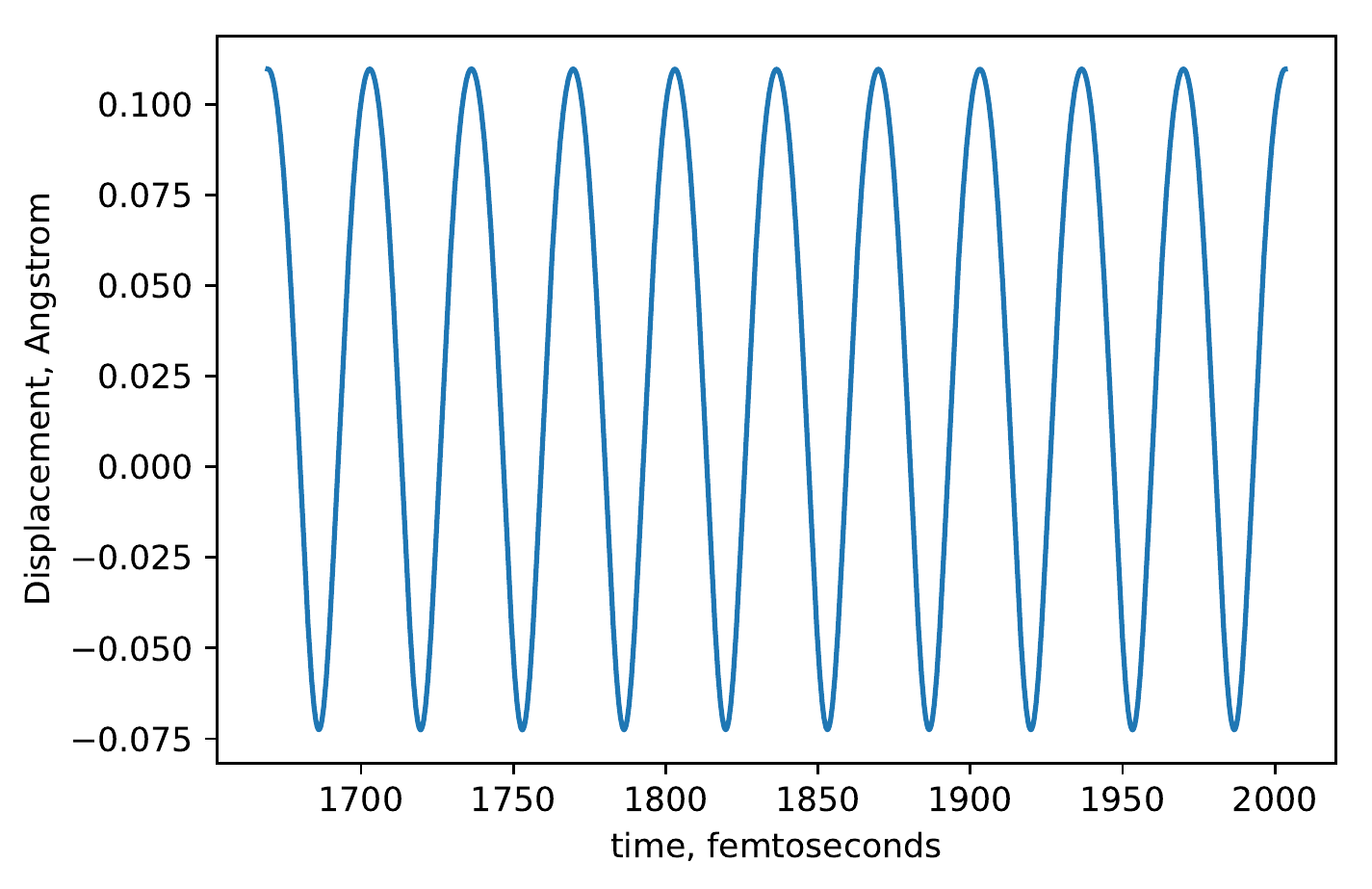} \\ (b)}
	\end{minipage}
	\hfill
	\begin{minipage}[h]{0.32\linewidth}
		\center{\includegraphics[width=1\linewidth]{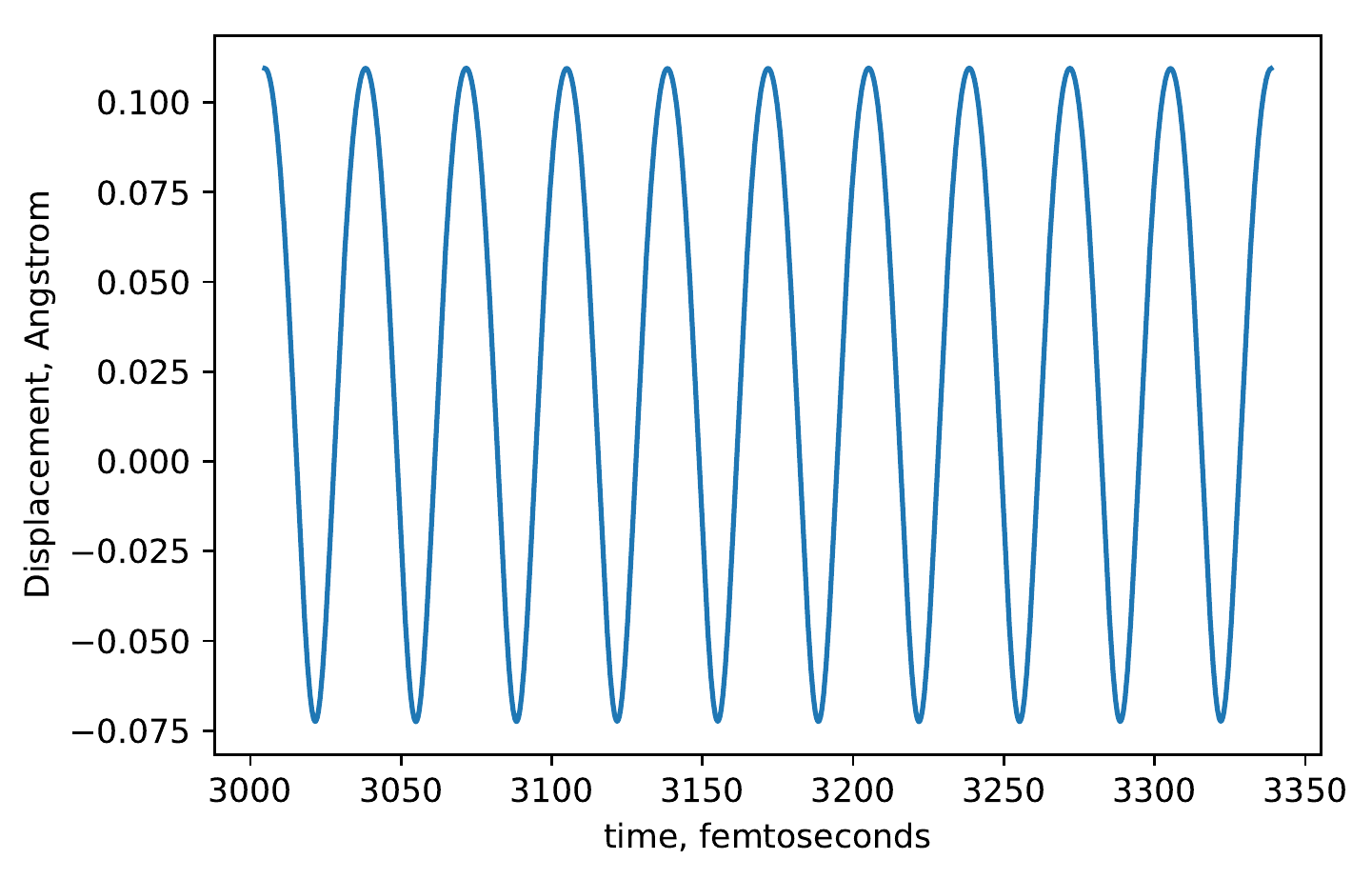} \\ (c)}
	\end{minipage}
	\caption{Atomic oscillations, corresponding to mode~2 with amplitude $0.11$~\AA\ for three different subintervals: (a)~$0\isep 10T$; (b)~$50T\isep 60T$; (c)~$90T\isep 100T$, where $T$ is the period of oscillations.}
	\label{fig9}
\end{figure}

\begin{figure}[h]
	\centerline{\includegraphics[width=0.5\linewidth]{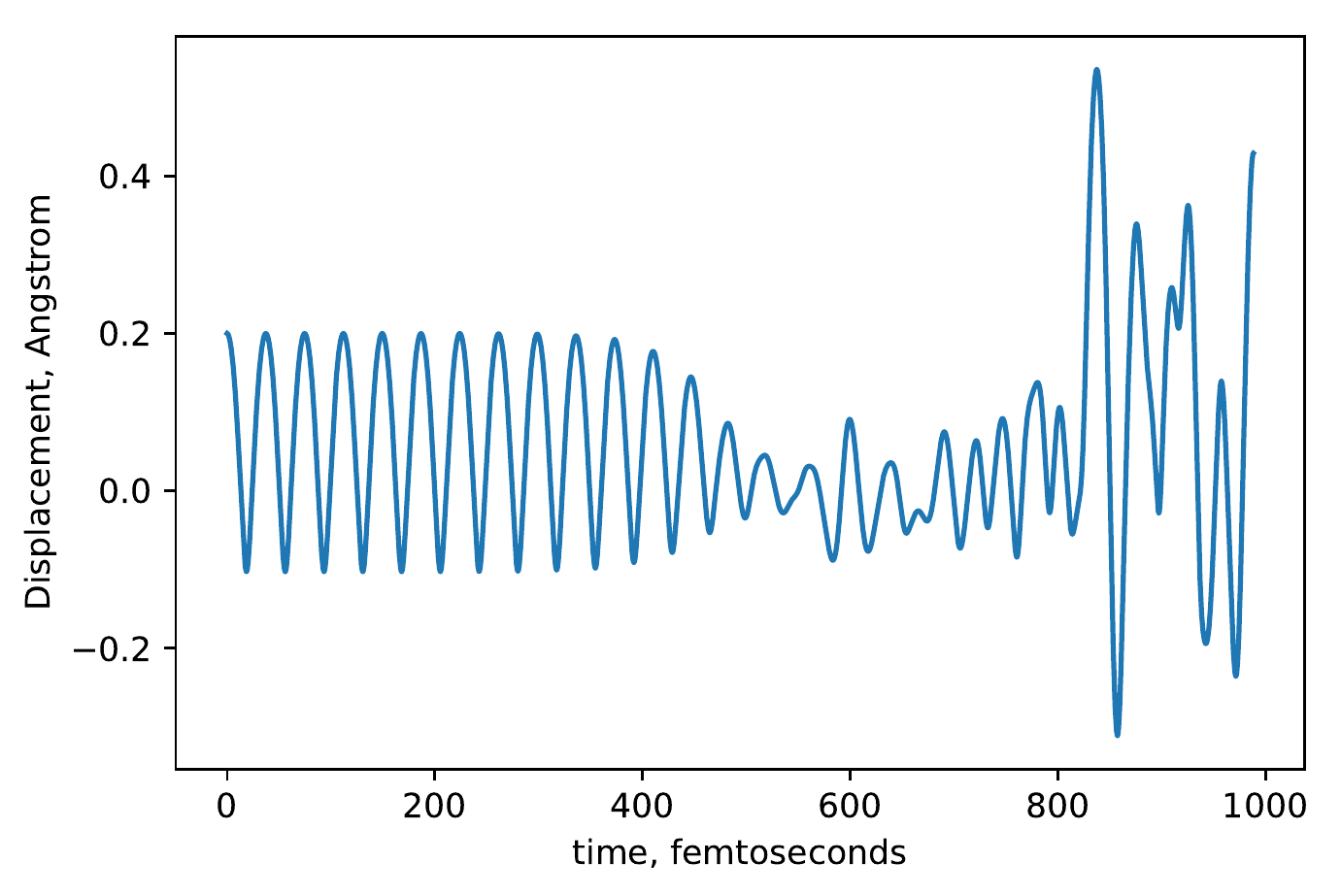}}
	\vspace*{3pt}
	\caption{Atomic oscillations, corresponding to mode~2 with amplitude $0.2$~\AA.}
	\label{fig10}
\end{figure}

Let us now discuss dynamical properties of the mode~2.  This mode also demonstrates a soft type of nonlinearity. Indeed, its frequency decreases from $31.35$ THz to $27.78$ THz when the amplitude increases from $0.05$~\AA\ to $0.3$~\AA. Frequency-amplitude characteristic of the mode~2 is shown in Fig.~\ref{fig6}b. Atomic oscillations, associated with this mode, are shown in Fig.~\ref{fig9} for sufficiently large time interval. In this figure one can see oscillations on three different subintervals of the total interval to demonstrate stability of the mode~2 with amplitude $0.11$~\AA. However, for the amplitude equal to $0.2$~\AA, this mode already loses stability as it follows from Fig.~\ref{fig10}.

Detailed studying of the stability of modes 1 and 2 for the diamond lattice will be published elsewhere.

\section{Conclusion}

In this paper, large-amplitude atomic vibrations in diamond were considered for the first time. We discussed two symmetry-determined nonlinear normal modes which can exist in the lattice of diamond structure, as well as group-theoretical methods of their construction. We also verify the correctness of these geometrical results with the aid of simulations on the base of the density functional theory. It occurs that these nonlinear normal modes demonstrate for the appropriate amplitudes stable periodic oscillations of all atoms of diamond crystal for sufficiently large time intervals. We plan to present a detail examination of the stability of these modes in a further paper.

We also discuss here the concept of the bushes of nonlinear normal modes that was developed in previous papers~\cite{DAN-1, DAN-2, PhysD-98}. They represent exact dynamical objects in nonlinear physical systems with discrete symmetry and describe, in general, a quasi-periodic atomic motion. In conclusion, let us note that these delocalized objects were used in~\cite{korznikova2017, barani2017} to construct discrete breathers, which represent time-periodic vibrations localized on the crystal lattice.

\section*{Acknowledgements}
\noindent
G.~Chechin acknowledges financial support by the Russians Science Foundation (Grant No.\ 14-13-00982); D.~Ryabov acknowledges financial support by the Ministry of Education and Science of the Russian Federation (state assignment grant No.\ 3.5710.2017/8.9). The authors are sincerely grateful to N.~Ter-Oganessian for useful discussions.

\bibliographystyle{ieeetr}
\bibliography{Biblio}

\begin{table}[h]
	\caption{Atomic displacements corresponding to mode~1 for all atoms in one primitive cell. Here $a$ is lattice constant, equal to $1.79$~\AA, while $\Delta$ is amplitude of the mode.}
	{\begin{tabular}{| c | c | c |}
			\hline
			No.~of atom & Atom position & $[x,y,z]$\\ & & displacements \\ \hline
			1 & $[\textover{0}{\frac12a},\textover{0}{\frac12a},\textover{0}{\frac12a}]$ & $[-\Delta,-\Delta,-\Delta]$ \\ \hline
			2 & $[\frac12a,\frac12a,\frac12a]$ & $[\hphantom{-}\Delta,\hphantom{-}\Delta,\hphantom{-}\Delta]$ \\ \hline
		\end{tabular} \label{table1}}
\end{table}
\begin{table}[h]
	\caption{Atomic displacements corresponding to mode~2 for all atoms in one Bravais-cell. Here $a$ is lattice constant, equal to $1.79$~\AA, while $\Delta$ is amplitude of the mode.}
	{\begin{tabular}{| c | c | c |}
			\hline
			No.~of atom & Atom position & $[x,y,z]$\\ & & displacements \\ \hline
			1 & $[\textover{0}{\frac12a},\textover{0}{\frac12a},\textover{0}{\frac12a}]$ & $[-\Delta,-\Delta,-\Delta]$ \\ \hline
			2 & $[\frac12a,\frac12a,\frac12a]$ & $[\hphantom{-}\Delta,\hphantom{-}\Delta,\hphantom{-}\Delta]$ \\ \hline
			3 & $[\textover{0}{\frac12a},\textover{a}{\frac12a},\textover{a}{\frac12a}]$ & $[\hphantom{-}\Delta,-\Delta,\hphantom{-}\Delta]$ \\ \hline
			4 & $[\frac12a,\frac32a,\frac32a]$ & $[-\Delta, -\Delta,  \hphantom{-}\Delta]$\\ \hline
			5 & $[\textover{a}{\frac12a},\textover{0}{\frac12a},\textover{a}{\frac12a}]$ & $[\hphantom{-}\Delta,\hphantom{-}\Delta,-\Delta]$\\ \hline
			6 & $[\frac32a,\frac12a,\frac32a]$ & $[\hphantom{-}\Delta,-\Delta,-\Delta]$\\ \hline
			7 & $[\textover{a}{\frac12a},\textover{a}{\frac12a},\textover{0}{\frac12a}]$ & $[-\Delta,\hphantom{-}\Delta,\hphantom{-}\Delta]$\\ \hline
			8 & $[\frac32a,\frac32a,\frac12a]$&  $[-\Delta,\hphantom{-}\Delta,-\Delta]$\\ \hline
		\end{tabular} \label{table2}}
\end{table}

\end{document}